\documentclass[a4paper,fleqn,usenatbib]{mnras}

\usepackage[T1]{fontenc}
\usepackage{ae,aecompl}

\usepackage{graphicx}	
\usepackage{amsmath}	
\usepackage{amssymb}	

\title[Lyman photon escape in local starburst galaxies]{Investigating the Lyman photon escape in local starburst galaxies with the Cosmic Origins Spectrograph
\thanks{Based on observations made with the Hubble Space Telescope under program ID: 13313 }}

\author[S. Hernandez et al.]
{\parbox{\textwidth}{Svea Hernandez,$^{1}$\thanks{E-mail: s.hernandez@astro.ru.nl}
Claus Leitherer,$^{2}$
M\'{e}d\'{e}ric Boquien,$^{3}$
V\'{e}ronique Buat,$^{4}$
Denis Burgarella,$^{4}$
Daniela Calzetti,$^{5}$
and Stefan Noll $^{6,7,8}$
}
\\ \\
$^{1}$Department of Astrophysics / IMAPP, Radboud University, PO Box 9010, 6500 GL Nijmegen, The Netherlands\\
$^{2}$Space Telescope Science Institute, 3700 San Martin Drive, Baltimore, MD 21218, USA\\
$^{3}$Universidad de Antofagasta, Avenida Angamos 601, Antofagasta 1270300, Chile\\
$^{4}$Aix-Marseille Universit\'{e}, Laboratoire d'Astrophysique de Marseille, UMR7326, 13388, Marseille, France\\
$^{5}$Department of Astronomy, University of Massachusetts-Amherst, Amherst, MA 01003, USA\\
$^{6}$Institut f\"ur Physik, Universit\"at Augsburg, 86135 Augsburg, Germany\\
$^{7}$German Aerospace Center, German Remote Sensing Data Center, Oberpfaffenhofen, 82234 We{\ss}ling, Germany\\
$^{8}$Institute f\"ur Astro- und Teilchenphysik, Universit\"at Innsbruck, Technikerstr. 25/8, 6020 Innsbruck, Austria\\
}

\date{Accepted 2018 April 24. Received 2018 April 24; in original form 2017 December 14}

\pubyear{2017}

\begin{document}
\label{firstpage}
\pagerange{\pageref{firstpage}--\pageref{lastpage}}
\maketitle

\begin{abstract}
We present a study of 7 star-forming galaxies from the Cosmic Evolution Survey (COSMOS) observed with the Cosmic Origins Spectrograph (COS) on board the Hubble Space Telescope (HST). The galaxies are located at relatively low redshifts, $z\sim$0.3, with morphologies ranging from extended and disturbed to compact and smooth. To complement the HST observations we also analyze observations taken with the VIMOS spectrograph on the Very Large Telescope (VLT). In our galaxy sample we identify three objects with double peak Lyman-$\alpha$ profiles similar to those seen in Green Pea compact galaxies and measure peak separations of 655, 374, and 275 km s$^{-1}$. We measure Lyman-$\alpha$ escape fractions with values ranging between 5-13\%. Given the low flux levels in the individual COS exposures we apply a weighted stacking approach to obtain a single spectrum. From this COS combined spectrum we infer upper limits for the absolute and relative Lyman continuum escape fractions of $f_{\rm abs}(\rm LyC)$ = 0.4$^{+10.1}_{-0.4}$\% and $f_{\rm res}(\rm LyC)$ = 1.7$^{+15.2}_{-1.7}$\%, respectively. Finally, we find that most of these galaxies have moderate UV and optical SFRs (SFRs $\lesssim$ 10 M$_{\odot}$ yr$^{-1}$). 
\end{abstract}




\section{Introduction}
Star formation is the fundamental process transforming baryonic matter in the Universe, converting the hydrogen reservoir into heavy elements. Consequently star formation is one of the main drivers of galaxy formation and evolution. Tracing accurately star formation is therefore of critical importance for understanding the fundamental processes operating in galaxies across cosmic time \citep{ken12,mad14}.\par

The global stellar spectral energy distribution (SED) of star-forming galaxies is dominated by radiation of massive stars with masses of tens of solar masses (M$_{\odot}$). Such stars emit the bulk of their energy in the ultraviolet (UV). Typical star-forming galaxies emit approximately 75\% of their stellar radiation in the Balmer continuum between 912 and 3646 \r{A}, with an additional $\sim$25\% contributed by the Lyman continuum \citep{lei02}. The intrinsic stellar energy output longward of the Balmer break is typically negligible. Notably, the wavelength ranges below and above Lyman-$\alpha$ (Ly$\alpha$) at 1216 \r{A} make roughly equal contributions to the Balmer continuum: the energy radiated accounts for about 40\% of the bolometric luminosity of the full stellar SED. Characterizing the behavior of the UV wavelength region below Ly$\alpha$, above and below the Lyman break is therefore essential for understanding the recent star formation in galaxies.\par

Available spectroscopic data for the wavelength range below 1216 \r{A} in the observed frame are scarce. Among the previous space missions exploring this wavelength domain only the Hopkins Ultraviolet Telescope \citep[HUT, ][]{dav92, kru99} and the Far Ultraviolet Spectroscopic Explorer \citep[FUSE, ][]{moo00} had the sensitivity necessary for significant extragalactic studies. More recently, the Hubble Space Telescope (HST) Cosmic Origins Spectrograph \citep[COS, ][]{ost11, gre12} has been added to the list of instruments with spectroscopic capabilities down to and below 912 \r{A}. COS has become the instrument of choice for investigating the properties of star-forming galaxies in the local universe below the Lyman limit in particular. \par 

Understanding the transformation of the universe from neutral and opaque during the Dark Ages to transparent after the reionization has been an increasingly important topic in observational cosmology. There is still great uncertainty on the population of objects that reionized the Universe. Candidates range from primordial black holes and mini quasars \citep{mad04} to active galactic nuclei \citep[AGN, ][]{hai98,mad15}, and star-forming galaxies \citep{rob10}. Although AGNs have been observationally confirmed to exist at the beginning of the epoch of reionization \citep{ven07}, their population numbers may be insufficient to contribute significantly to the cosmic reionization. Star-forming galaxies are more commonly accepted as strong contributors of the ionizing background at high redshift \citep[z$>$6, ][]{rob10}, however, the sources identified so far are insufficient to complete the ionization of the Universe by z $\sim$ 6 \citep{iwa09, cow09, rob13}. At high-z the observed UV luminosity function appears to be steep, therefore faint, low-mass star-forming galaxies could in principle be responsible for most of the ionizing radiation \citep{ouc09, yaj11, bou15}. \par

The presence of ionizing stars in star-forming galaxies has been inferred at z $\lesssim$ 7 using a variety of proxies such as dust emission, stellar absorption lines, nebular emission lines, and blue galaxy colors \citep{bou12, bar12, sta13, fin16}. With these confirmed sources of ionizing photons at relatively high-z, measurements of the Lyman continuum escape fraction, $f_{\rm esc}(\rm LyC)$, are critical for identifying the conditions and environments that benefit their escape onto the intergalactic medium (IGM). Extensive observational studies covering a broad range of redshifts have been done to detect escaping Lyman continuum photons from star-forming galaxies \citep{ber13}. It has been proposed that for galaxies to completely reionize the intergalactic \ion{H}{I}, their Lyman continuum escape fractions need to be on the order of 10-20\% \citep{rob13}. \par
Efforts to observe escaping Lyman continuum radiation at relatively high redshifts (z $>$ 1) are challenging not only because of possible contamination from low-redshift interlopers \citep{van10,mos15}, but at redshifts z $>$ 3 these star-forming galaxies are significantly affected by attenuation by the Ly$\alpha$ forest along the line of sight \citep{ino14}. In spite of these complications, detections of escaping Lyman continuum radiation have been suggested in $\sim$10\% of surveyed galaxies \citep{sia15}.\par 
Attempts to measure the $f_{\rm esc}(\rm LyC)$ of local galaxies were made using HUT and FUSE. Various upper limits were established \citep{lei95, hec01, gri09}, and only a small number of weak detections of $f_{\rm esc}(\rm LyC)$ were found \citep{lei11, lei13}. In recent years COS has kept pushing the limits of this field by providing direct detections of $f_{\rm esc}(\rm LyC)$ in starburst galaxies at low redshifts with values as high as 43\% \citep{bor14, lei16, izo16a, izo16b, izo18}.\par

Similar to Lyman continuum photons, Ly$\alpha$ photons originate in young star forming galaxies, but these instead come from the recombination of hydrogen gas. In the last couple of decades many Ly$\alpha$ emission line (LAE) galaxies have been observed \citep{hu98, rho00, gaw06, wan09, kas11, mat14, zhe16}. Since these Ly$\alpha$ photons propagate throughout the IGM and scatter away by \ion{H}{i}, the Ly$\alpha$ emission line can be used to study the reionization of the IGM \citep{mal04, til14, mat15}. In this context, the Ly$\alpha$ escape fraction represents the number of Ly$\alpha$ photons that escape the interstellar medium of their LAEs. A large fraction of Lyman continuum leakers have been found to be Ly$\alpha$ emitters \citep{lei13, bor14, izo16a, izo16b, lei16, sha16, izo18}. For this reason, several studies have proposed to use Ly$\alpha$ profiles as tools for identifying Lyman continuum leakers \citep{ver15, ver17, dij16}.

In this work we take advantage of the far-UV capabilities of COS and analyze spectra of low redshift galaxies, that we observed in Cycle 21. Our main goals are: (i) probing the star-formation properties of normal galaxies at wavelengths where the stellar populations emit the majority of their energy; (ii) measuring or providing upper limits to the escape fractions of Lyman continuum and Ly$\alpha$ photons; (iii) comparing several panchromatic star-formation tracers. This paper is structured as follows. In Sections ~\ref{sec:targ_sel} and ~\ref{sec:obs_data} we present our target selection and HST observations/analysis, respectively. We discuss the spectral morphology of our objects in Section ~\ref{sec:spectra}. In Sections ~\ref{sec:lyalpha} and ~\ref{sec:SF} we describe the Ly$\alpha$ profiles and star formation properties, correspondingly. The Lyman continuum discussion is included in Section ~\ref{sec:esc}. Finally, in Section ~\ref{sec:conclusion} we present our conclusions. \par

\section{Target Selection} \label{sec:targ_sel}
We exploit the data from HST program 13313 (PI: Boquien) observing 8 star-forming galaxies with COS onboard HST. The program observed galaxies at redshifts around $z\sim0.3$ given the multiple advantages of this redshift i.e. the Lyman break is redshifted into the COS FUV sensitive range, and spiral galaxies at these distances have angular sizes comparable to the COS primary science aperture (PSA).   \par
The primary selection criteria in Program 13313 were: (i) The galaxies have a redshift $z >$ 0.25 to ensure that the UV domain is observed down to the Lyman break. (ii) The galaxies do not have any detected AGN activity to prevent any contamination which could affect the SED. (iii) The galaxies have a size small enough so that the bulk of the FUV emission is enclosed in the COS aperture. (iv) The galaxies are brighter than 22.5 AB magnitudes in the FUV within the COS aperture to ensure high enough signal-to-noise ratios (S/N) in a reasonable exposure time. The galaxy sample was selected from the Cosmic Evolution Survey \citep[COSMOS, ][]{sco07} with its rich body of ancillary spectroscopic and photometric data including Herschel fluxes from 70 to 500 $\micron$. After selecting all galaxies fulfilling the aforementioned criteria, the team constituted a sample of eight galaxies spanning FUV attenuation values of 0.8 $\leq$ A$_{\rm FUV}$ $\leq$ 2.5.
\par
In Table ~\ref{tab:prog} we present some information for these galaxies: target identifications (IDs) as listed in MAST, coordinates, Galaxy Evolution Explorer \citep[GALEX, ][]{mar05} FUV flux, Galactic foreground extinction, redshift, and distance ($D$). The GALEX FUV flux were obtained by extracting the GALEX FUV magnitudes from the public COSMOS catalog, and converting these to fluxes. The Galactic foreground extinction is adopted from the NASA Extragalactic Database (NED) which is based on the \citet{sch11} recalibration of the extinction maps by \citet{sch98}. Additionally, the coordinates listed in Table ~\ref{tab:prog} are based on the COSMOS HST ACS \textit{I} band survey \citep{san07}. The luminosity distance values, $D$, are obtained with the cosmological calculator \citep[NED, ][]{wri06} using the parameters from \citet{pla15}: H$_{0}$ = 67.8 km s$^{-1}$ Mpc$^{-1}$ and $\Omega_{m} = 0.308$ for a flat universe where $\Omega_{\rm vac} = 1 - \Omega_{m}$. At a redshift of 0.25 $\leq z \leq$ 0.32, the star-forming regions of these galaxies are enclosed by the 2.5$\arcsec\;$ COS aperture while the instrument provides a coverage down below the rest-frame Lyman break at such redshifts.\par

The sample galaxies in program 13313 are moderately star-forming objects, with specific star formation rates (sSFRs) ranging between $\log({\rm sSFR}) = -9.6$ yr$^{-1}$ and  $\log({\rm sSFR}) = -8.4$ yr$^{-1}$, making these galaxies hitherto unexplored yet difficult to observe. With exposure times $\sim$ 5000s we reach on average fluxes of $\sim$3$\times$ 10$^{-17}$ erg s$^{-1}$ cm$^{-2}$ \r{A}$^{-1}$ at 1100 \r{A} (rest wavelength). \par
\begin{figure*}
\centering
\includegraphics[scale=0.22]{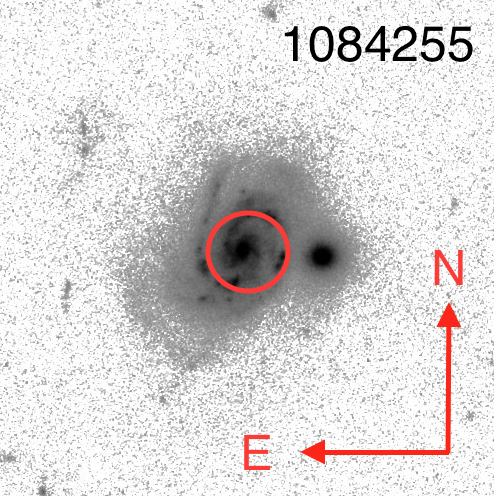}
\includegraphics[scale=0.22]{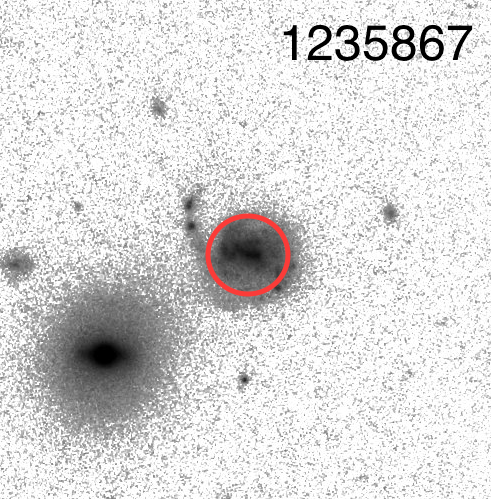}
\includegraphics[scale=0.22]{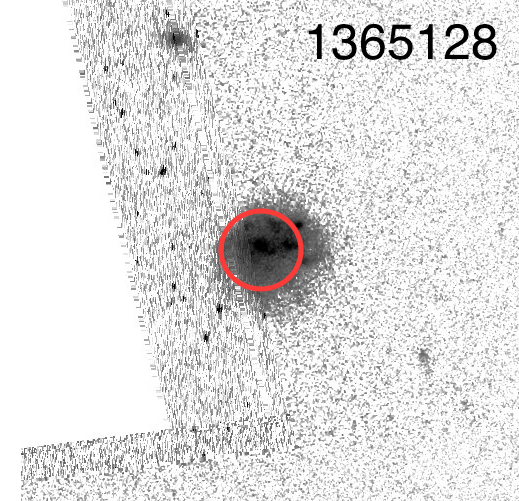}
\includegraphics[scale=0.22]{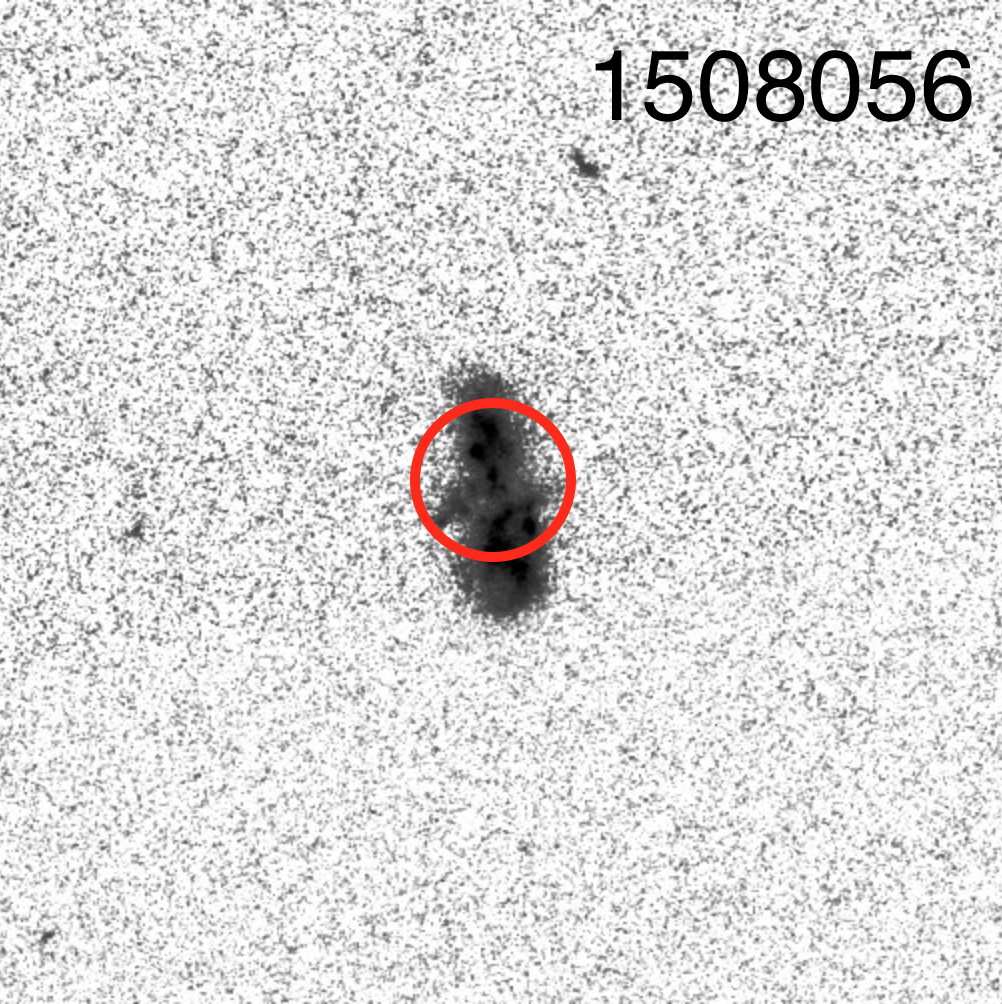}
\includegraphics[scale=0.22]{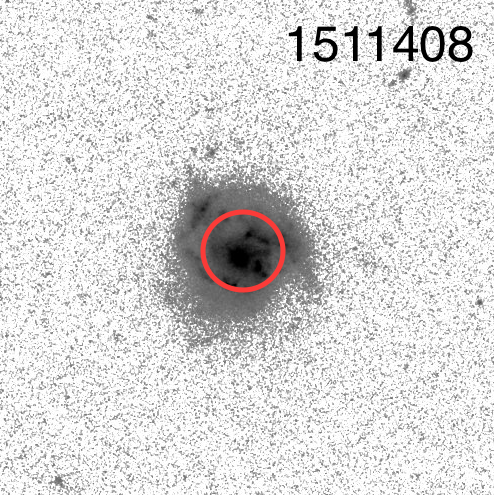}
\includegraphics[scale=0.22]{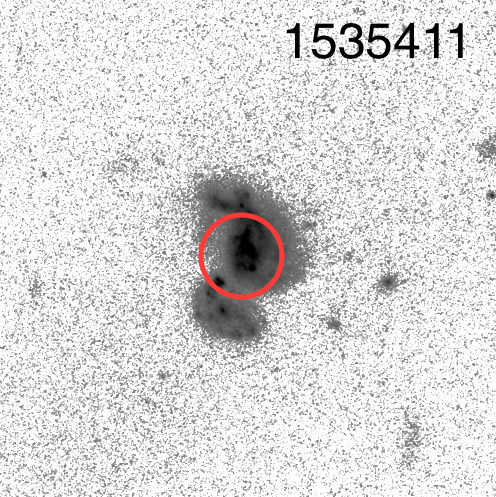}	
\includegraphics[scale=0.22]{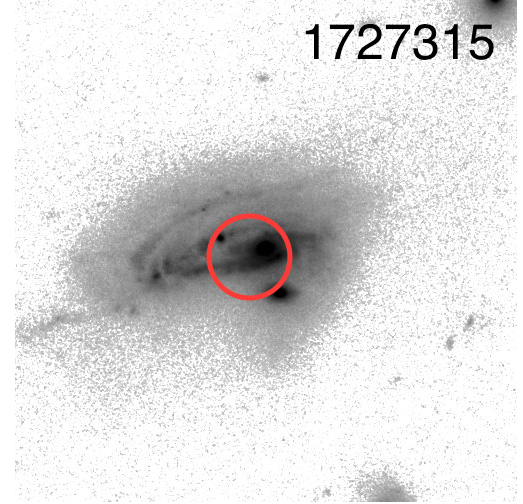}
\includegraphics[scale=0.22]{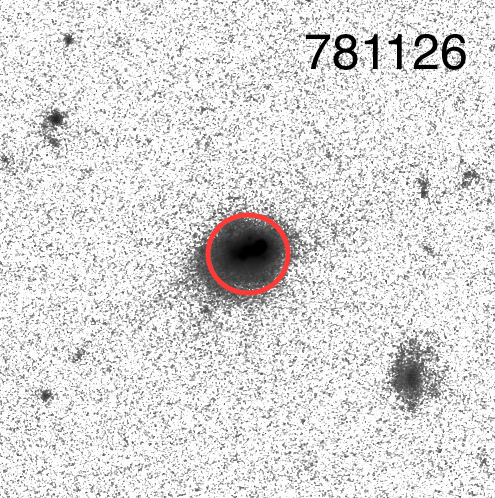}
\caption{ 15$\arcsec$ $\times$ 15$\arcsec$ HST/ACS Mosaic $I$-band (F814W) from COSMOS for each of the galaxies studied in this work. The red circle shows the COS 2.5\arcsec  aperture. Some uncertainty is expected in the final pointing, of the order of $\sim0.3\arcsec \:$ RMS, as the objects were observed using the blind pointing mode. Each stamp has north up, and east to the left as shown in the first panel.}
\label{fig:cosmos}
\end{figure*}

In Figure ~\ref{fig:cosmos} we show the selected galaxies as observed in COSMOS. The different panels present 15\arcsec\: $\times$ 15\arcsec\: Advanced Camera for Surveys (ACS) Mosaics as seen in the $I$-band (F814W) along with red  circles showing the COS spectroscopic 2.5\arcsec\: aperture. The panels show the diverse morphologies of the sample galaxies ranging from extended and disturbed, i.e. 1727315, 1535411 and 1084255, to compact and smooth, i.e. 1365128, 1508056, and 781126.

\begin{table*}
\caption{Program Galaxies} 
\label{tab:prog}
\centering
\begin{tabular}{ccccccccc}
 \hline  \hline \
Target & RA (J2000) & Dec (J2000) & GALEX FUV Flux& $E(B-V)_{\rm MW}$ & $z^a$ & D$^b$\\
& (h m s) & ($^{\circ}$ $\arcmin$ $\arcsec$) &  (10$^{-16}$ erg sec$^{-1}$ cm$^{-2}$ \r{A}$^{-1}$) & (mag) & & (Mpc)\\
  \hline \\ 
1084255& 09 \:58\: 15.487&	+02 \:11 \:35.50 & 2.87&0.017& 0.265&	1388\\
1535411&	09 \:58\: 44.093&	+02 \:28 \:43.97 & 2.32&0.018& 0.314& 1686\\
1511408&	09 \:59\: 03.605&	+02 \:28 \:33.19 & 2.10&0.017& 0.261& 1364\\
1508056&	09 \:59\: 08.734&	+02\: 30 \:29.56 & 1.02&0.016& 0.320& 1724\\
1727315&	09 \:59\: 21.341&	+02 \:40 \:30.29 & 2.74&0.017& 0.260& 1358\\
1235867&	10 \:00\: 22.166&	+02 \:21 \:41.26& 1.28&0.015& 0.265& 1388\\
781126& 10 \:00\: 35.726&	+02 \:01 \:13.43 & 3.02&0.016& 0.267& 1400\\
1365128&	10 \:02\: 55.675&	+02 \:30 \:25.34 & 1.53&0.018& 0.269& 1412\\
 \hline 
 \end{tabular}
\begin{minipage}{12cm}~\\
  \textsuperscript{$a$}{Redshifts are extracted from the public COSMOS catalog.}\\
 \textsuperscript{$b$}{Luminosity distance.}
 \end{minipage}
\end{table*}

\section{HST/COS Observations and Data Reduction} \label{sec:obs_data}
The selected galaxies were observed with COS onboard HST \citep{gre12} between March and May of 2014.  After the guide stars were acquired, the targets were observed in blind pointing mode. No target acquisitions were requested for the observations as the standard HST pointing accuracy ($\sim$0.3\arcsec) was deemed sufficient for the program. We discuss flux losses due to pointing offsets in Appendix \ref{appendix_nuv}. For each orbit a short exposure was taken in ACCUM mode using the MIRRORA (imaging mode) element. The science observations were taken in TIME-TAG mode using the low-resolution G140L grating ($R\sim2000$)  configured to observe at central wavelength 1105 \r{A}. Since the COS G140L/1105 setting shifts the zero-order image to detector Segment B, only Segment A is available for this configuration. G140L/1105 provides data with a wavelength coverage between 1105  to 2250 \r{A} (observed frame). Originally the FUV channel was designed to operate optimally between wavelengths 1130 $<$ $\lambda$ $<$ 1850 \r{A}. At wavelengths $<$ 1130 \r{A} the sensitivity declines by a factor of 100$\times$, but is still comparable to that of FUSE \citep{mcc10}. To reduce the well-known fixed pattern noise we used all four focal plane positions (FP-POS). In Table \ref{tab:obslog} we  provide information on the individual exposures including dataset names, start times, total duration and S/N. The S/N values per resolution element are low as the original HST program relies on binned spectra.  \par
Before proceeding with the calibration and analysis we inspect the short-exposure images taken before the science observations and confirm that the galaxies are indeed on the detector. We present a brief discussion of these images in Appendix \ref{appendix_nuv}. We retrieve the individual datasets from MAST and calibrate the exposures using a modified version of the official CalCOS pipeline introduced in \citet{lei16}. The standard CalCOS v3.2.1 software \citep{fox15} performs a 1D spectral background subtraction, first computing the number of counts in predefined regions external to the science extraction box, scaling these counts by the ratio of pixel heights of the science/background regions, and finally subtracting these scaled counts from the science spectrum at each wavelength. The actual background at the target location might differ from that found in the predefined background regions, a behavior that might cause CalCOS v3.2.1 to over- or under-subtract the background contribution from the final calibrated product. \par
As mentioned earlier, for wavelengths  $<$1130 \r{A} the sensitivity declines by a factor of 100$\times$, therefore for observations taken in the background-limited regime an optimal detector background subtraction is critical. In order to perform a more accurate and tailored background correction in \citet{lei16} we present a modified version of an older CalCOS version (v2.21) where we introduce a two-dimensional (2D) background correction applicable to data obtained prior to 2015 February. For the analysis of these 8 galaxies, we further  modify CalCOS v2.21d to improve the handling of negative values obtained after the 2D-superdark subtraction. More details on the exact changes made to CalCOS v2.21d are described in the Appendix \ref{appendix_calcos}. 
In addition to the optimized background correction, we also perform a dedicated analysis of the COS pulse-height amplitude (PHA\footnote{The PHA characterizes the total charge in the electron cloud incident triggered by an incoming photon. A PHA value basically defines the size of the electron shower/cloud.}) similar to that described in \citet{lei16}. The modifications to the PHA filtering do not introduce additional noise. And finally, to further increase the S/N we decrease the size of the spectral extraction box from the standard 57-pixel size to a 20-pixel size as suggested by the COS Team\footnote{\href{http://www.stsci.edu/hst/cos/documents/newsletters/cos_stis_newsletters/full_stories/2015_03/bkg_limited_targets}{http://www.stsci.edu/hst/cos/documents/newsletters/
cos\_stis\_newsletters/full\_stories/2015\_03/bkg\_limited\_targets}}.\par
The calibrated x1d files are taken and further analyzed using the IDL software developed by the COS Guaranteed Time Observer (GTO) Team \citep{dan10}. This software weight combines the different FP-POS exposures by interpolating onto a common wavelength vector accounting for the non-Poissonian noise as described by \citet{kee12}. We opt for a standard weighting scheme where the weights are defined as $w = 1/\sigma_{i}^{2}$, and $\sigma_{i}$ are the individual errors as extracted from the calibrated x1d files. We bin the data by a COS resolution element (1 resel = 6 pixels) which corresponds to the nominal point-spread function.

\begin{table*}
\caption{Observation Log} 
\label{tab:obslog}
\centering
\begin{tabular}{ccccc}
 \hline  \hline \
Target & Dataset & Start Time & Total Duration & S/N  \\
&  & (UT) & (s) & (resel$^{-1}$)$^{a,b}$  \\
  \hline\\ 
1084255& LC8802020&	2014-03-31 10:57:34 &4889.600& 0.3\\
& LC8802040& 	2014-03-31 12:29:54 & &	\\
1535411& LC8807020&	2014-05-21 08:57:37	&4889.760& 0.4\\
&  LC8807040	&2014-05-21 10:29:27 &&	\\
1511408& LC8806020&	2014-05-29 00:08:50	 &4889.696& 1.0\\
& LC8806040&	2014-05-29 01:41:31 &	&	\\
1508056&LC8805020&	2014-05-10 06:52:26 & 4889.760& 0.7	\\
&	LC8805040	&2014-05-10 08:24:39 &&	\\
1727315$^*$&LC8808020&	2014-03-24 14:49:53 &4889.664& 1.4	\\
&	LC8808040	 &2014-03-24 16:21:27 &&\\
1235867&  LC8803020&	2014-04-15 09:26:06 &4889.632&  0.8	\\
& LC8803040	&2014-04-15 10:58:17 &&\\
781126& LC8801020&	  2014-03-31 07:46:29 & 4889.600 & 1.8\\
& LC8801040&	  2014-03-31 09:18:50 &&\\
1365128 & LC8804020&	2014-04-16 02:57:59 &4889.664 &0.9	\\
&  LC8804040&	2014-04-16 04:30:06 &&	\\
\hline
\end{tabular}
\begin{minipage}{10cm}~\\
 \textsuperscript{$a$}{ 1 resel = 6 pixels }\\
 \textsuperscript{$b$}{ S/N at $\lambda_{obs}$=1250 \r{A} }\\
 \textsuperscript{$*$}{ This target is classified as a QSO in SDSS and therefore excluded from the analysis.}
\end{minipage}
\end{table*}

\section{Spectral morphology} \label{sec:spectra}
\begin{figure}
   \centering
            {\includegraphics[scale=0.6]{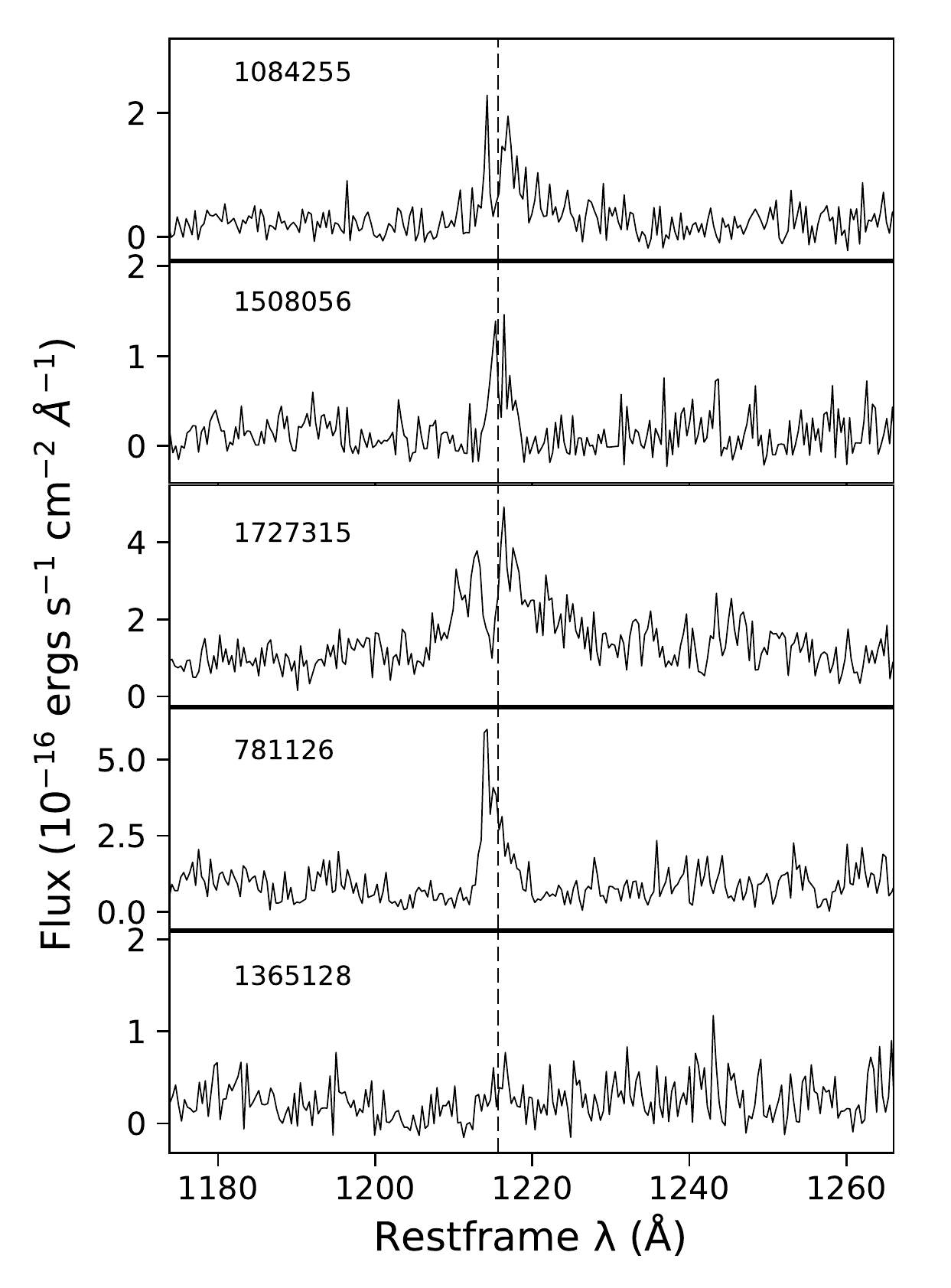}}
      \caption{Ly$\alpha$ emission lines present in the galaxy sample. The observed spectra have been redshift and foreground corrected. The vertical dashed line marks the location of the Ly$\alpha$ line.}
         \label{fig:lyman}
   \end{figure}

Before addressing the star formation properties of the individual galaxies we correct the spectra for Galactic foreground reddening using the $E(B-V)_{\rm MW}$ values shown in Table ~\ref{tab:prog} along with the reddening law of \citet{mat90}. We also transform the wavelength array from the observed to the rest frame using the redshift values in Table ~\ref{tab:prog}. After some inspection, we notice that the Ly$\alpha$ line in 1727315 is remarkably wider than the rest of the lines (see Figure ~\ref{fig:lyman}). This same target is classified as a quasi-stellar object (QSO) in the Sloan Digital Sky Survey (SDSS) and thus excluded from the rest of our analysis.\par
\begin{figure*}
    \centering
            {\includegraphics[scale=0.4]{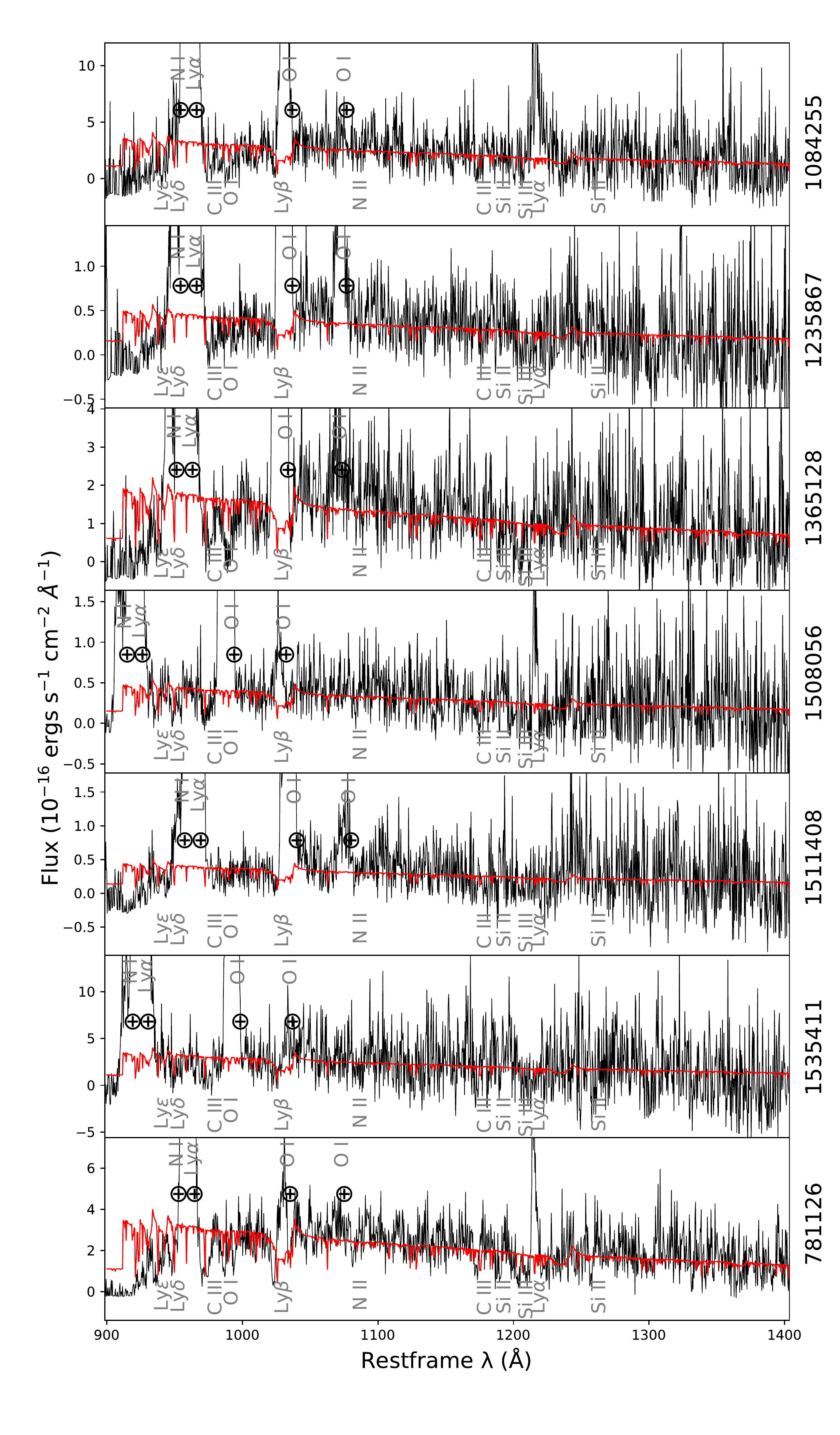}}
            {\includegraphics[scale=0.4]{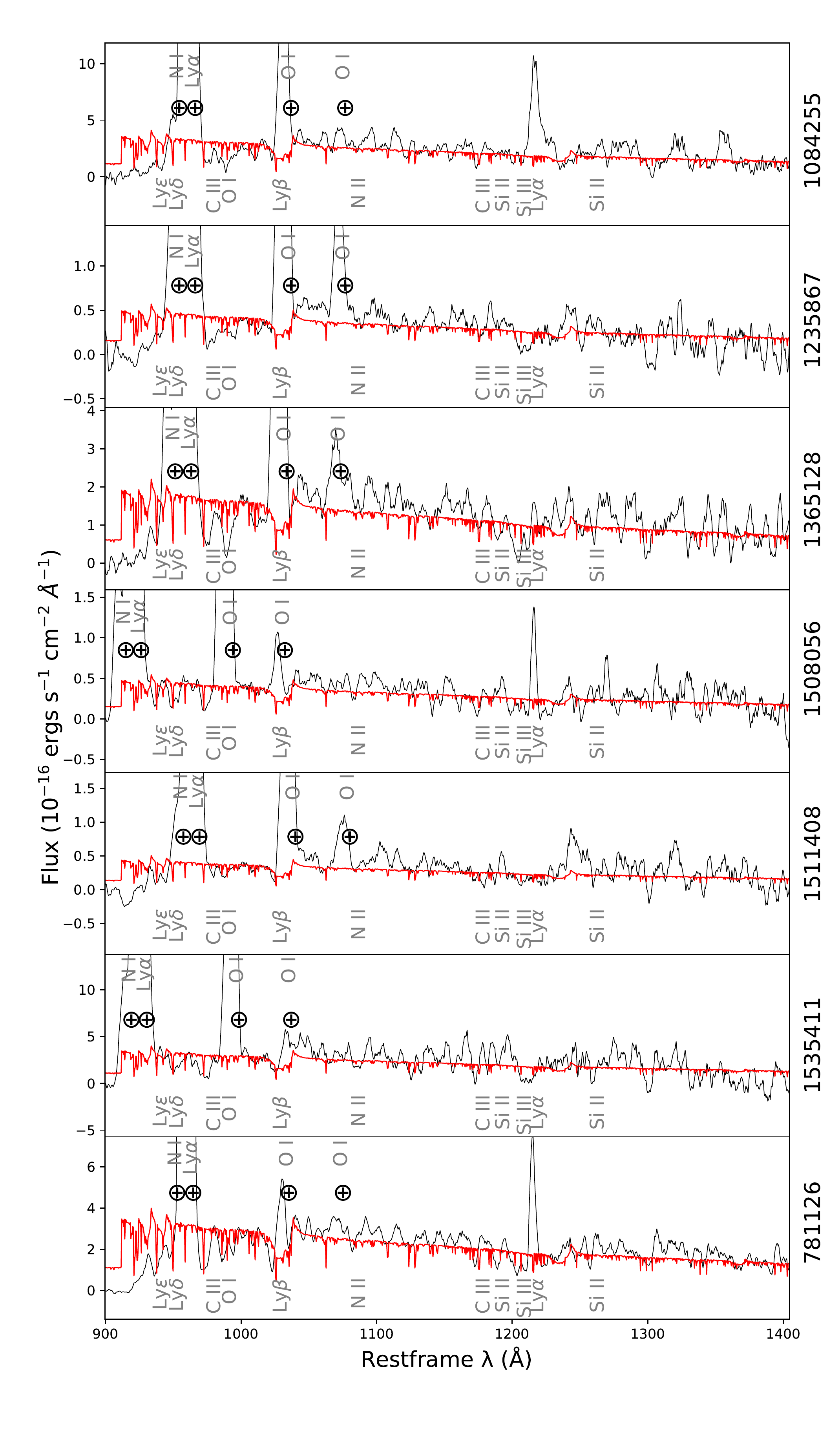}}
      \caption{Comparison of the observed spectra (black) with synthetic spectra of the best-fitting model (red). Left panel: Data binned by a COS resolution element. Right panel: For the benefit of visualization we smooth the observations using a boxcar size of 10 pixels. The observed spectra have been corrected for redshift, foreground and intrinsic reddening. The different SFRs are listed in Table ~\ref{tab:SF}. At the top and the bottom of the panels we identify the geocoronal ($\oplus$) and intrinsic spectral lines, respectively.  }
         \label{Fig:SpecALLMod}
   \end{figure*}

In Figure ~\ref{Fig:SpecALLMod} we show the galaxy spectra for each of the targets studied in detail here. We indicate the wavelengths of common spectral lines found in star-forming galaxies. The line identifications at the top of the figures refer to emission from geocoronal lines, whereas the bottom labels mark the absorption/emission from lines intrinsic to the galaxies.  The strongest emission observed in all of the targets come from geocoronal Ly$\alpha$ $\lambda_{\rm obs}1216$ and \ion{O}{I} $\lambda_{\rm obs}1302$.\par
We point out that not all the intrinsic lines marked in Figure ~\ref{Fig:SpecALLMod} are detected in every galaxy. The majority of the lines we detect are of interstellar origin, i.e. \ion{C}{III} $\lambda$977, \ion{O}{I}  $\lambda$988, and \ion{N}{II} $\lambda$1083. With both winds and ISM as expected mechanisms \citep{lei11} we observe \ion{C}{III} $\lambda$1175 most prominently in 781126. Additionally, we see Ly$\alpha$ $\lambda$1216  emission in 4 out of our 7 targets. We discuss in more detail the morphology of Ly$\alpha$ in the following section. \par

\section{Ly$\alpha$ emission} \label{sec:lyalpha}
The Ly$\alpha$ profiles for 1084255, 1365128, and 1508056 show double peak features resembling those observed in several Green Pea (GP) compact galaxies \citep{jas13,hen15,izo16b,izo18}. \citet{ver15} proposed that Ly$\alpha$ profiles with a double-peak morphology may indicate Lyman continuum leakage if the peak separation was $\lesssim300$ km s$^{-1}$. We measure peak separation values of 655, 374, and 275 km s$^{-1}$ for  1084255, 1365128, and 1508056, respectively. According to \citet{ver15}, 1084255 and 1365128 would be more unlikely to leak Lyman continuum photons given their high peak separation values.\par 

We point out that given that Ly$\alpha$ can be asymmetrically emitted and easily scattered, the Ly$\alpha$ radiation encompassed by the 2.5\arcsec\; COS aperture might not necessarily capture the Ly$\alpha$ emission fully. We estimate the Ly$\alpha$ escape fraction, $f_{\rm esc}$(Ly$\alpha$), by comparing the extinction-corrected Ly$\alpha$/H$\alpha$ flux ratios shown in Table ~\ref{tab:Lines}, and the intrinsic case B value of 8.7 corresponding to gas temperature of $T_{e} = 10000\: K$ and an electron density of $n_{e} = 350$ cm$^{-3}$ \citep[see][for a detailed discussion on adopting the 8.7 factor]{hen15}. The H$\alpha$ EWs and flux values are extracted from the VIMOS observations described in Section \ref{sec:optrange}. We measured the H$\alpha$ fluxes by fitting a Gaussian profile to the individual lines. On the other hand, given the complexity of the Ly$\alpha$ profiles we measure the EWs and fluxes using a simple flux over continuum integration code written in Python by \citet{pen13}. The continuum is estimated through a linear fit between wavelengths 1045--1550 \r{A}. The best fit is visually inspected to confirm the continuum placement is reasonable. The Ly$\alpha$ fluxes have been corrected for underlying stellar absorption using the model predictions by \citet{pen13}. We apply the recommended underlying correction for constant SFR over $\sim$20 Myr of the order of 7 \r{A}. We find $f_{\rm esc}$(Ly$\alpha$) in the range $\sim$ 5 - 13 $\%$. \par

\begin{figure}
\centering
            {\includegraphics[scale=0.43]{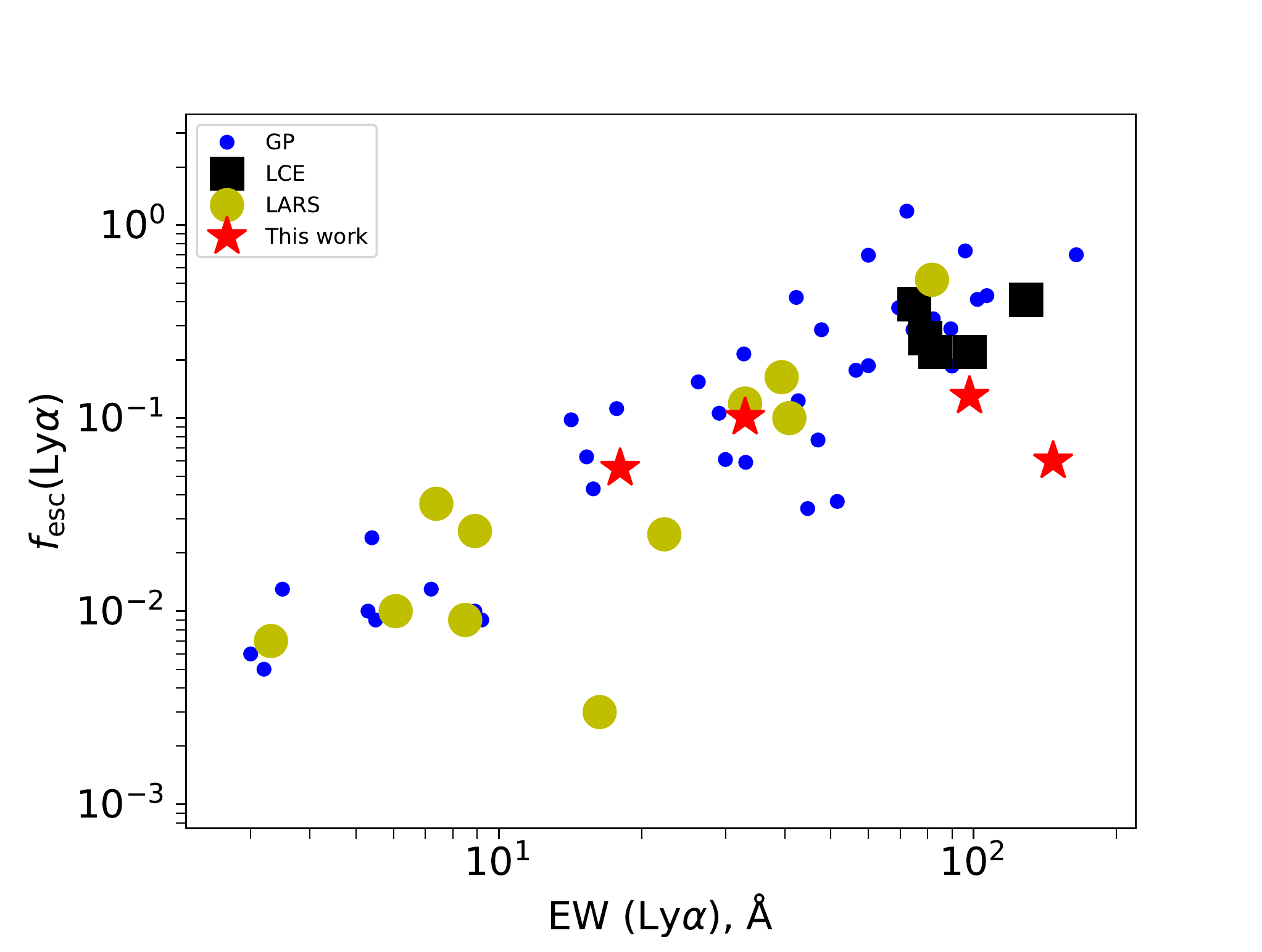}}
      \caption{In blue circles we show the $f_{\rm esc}$(Ly$\alpha$) as a function of EW(Ly$\alpha$) according to \citet{yan17}. Red stars show the values of 4 of the galaxies studied in this work. In yellow circles we show the escape fractions from \citet{hay14}. In black squares we display the measurements of \citet{izo16b}.}
         \label{fig:ly_fesc}
   \end{figure}

Several studies have shown that GP galaxies display strong Ly$\alpha$ emission \citep{jas14,hen15, ver17}. \citet{yan17} studied the Ly$\alpha$ profiles in a statistical sample of 43 GP galaxies and found that 2/3 of these galaxies are strong Ly$\alpha$ emitters. They also found a clear correlation between their Ly$\alpha$ EWs and the estimated $f_{\rm esc}$(Ly$\alpha$). Similar results have also been observed by \citet{ver17}. In Figure ~\ref{fig:ly_fesc} we show in blue circles the $f_{\rm esc}$(Ly$\alpha$) as a function of EW(Ly$\alpha$) as observed by \citet{yan17}. In this same figure we include our targets and their $f_{\rm esc}$(Ly$\alpha$) and EW(Ly$\alpha$) measurements as red stars, Lyman continuum emitters (LCEs) by \citet{izo16b} as black squares and Lyman Alpha Reference Survey galaxies \citep[LARS, ][]{hay13} by \citet{hay14} as yellow circles. In general we see that our galaxies follow the same $f_{\rm esc}$(Ly$\alpha$)-EW(Ly$\alpha$) trend as that observed by \citet{yan17} for GP galaxies, as well as that observed in LARS objects. We also find that 3/4 galaxies with Ly$\alpha$ emission in our sample have EW(Ly$\alpha$) $>$ 20 \r{A}, as pointed out by \citet{yan17} in a high-$z$ narrow-band study these objects would be classified as Ly$\alpha$ emitters.  \par

\begin{figure}
   \centering
            {\includegraphics[scale=0.5]{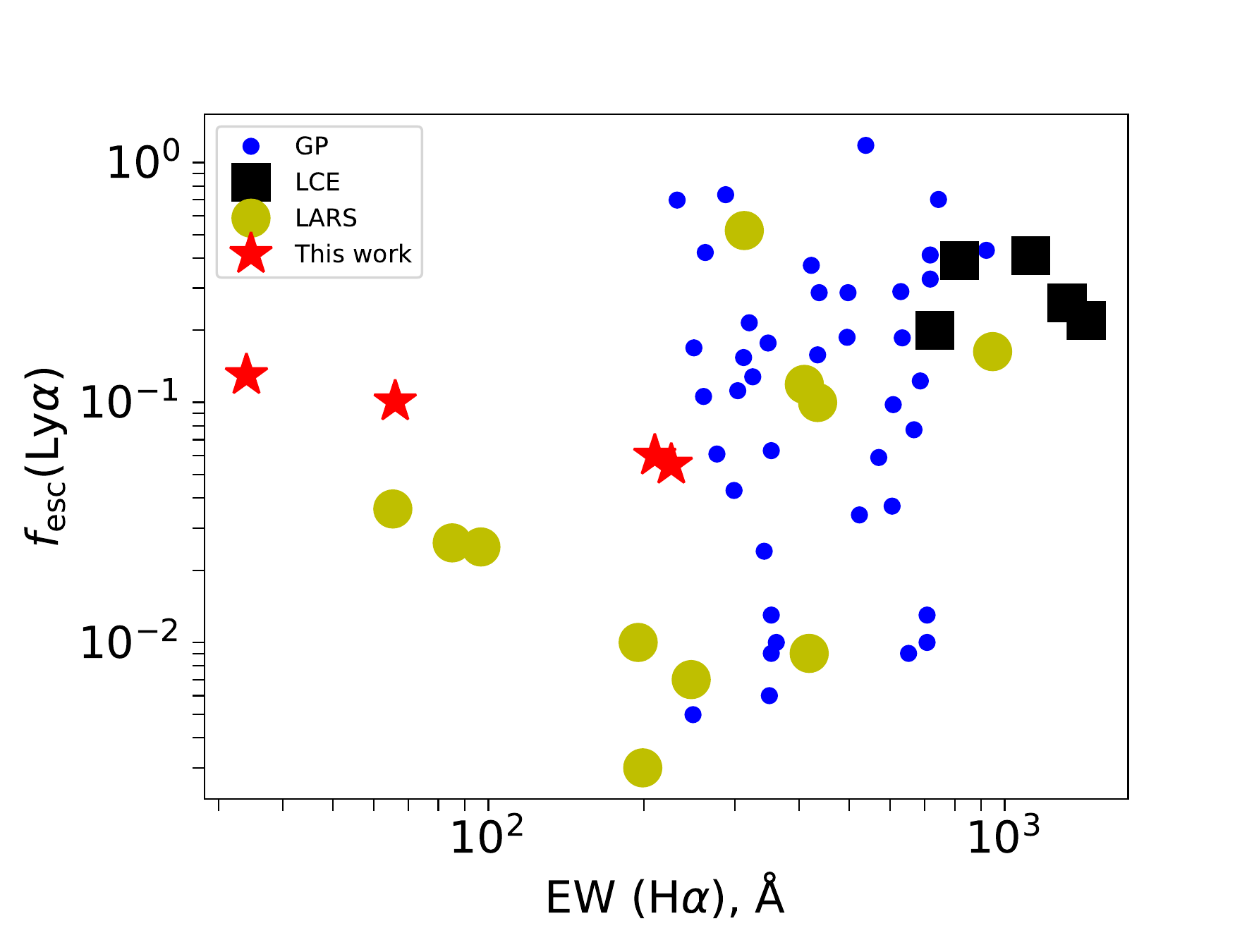}}
      \caption{$f_{\rm esc}$(Ly$\alpha$)  as a function of EW(H$\alpha$). In blue circle we show the measurements from \citet{yan17}. Red stars represent our measurements of star-forming galaxies. In yellow circles we show the escape fractions from \citet{hay14}. In black squares we display the measurements of \citet{izo16b}. }
         \label{fig:Halpha_fesc}
   \end{figure}

To put our star-forming galaxies in context we also compare our $f_{\rm esc}$(Ly$\alpha$) measurements to the dependence of $f_{\rm esc}$(Ly$\alpha$) on the EW(H$\alpha$) observed in GPs \citep{yan17}, LARS \citep{hay14}, and LCEs \citep{izo16b}. \citet{hay14} report global measurements of $f_{\rm esc}$(Ly$\alpha$) derived from Ly$\alpha$ and H$\alpha$ imaging of normal low redshift star-forming galaxies observed as part of LARS. From Figure ~\ref{fig:Halpha_fesc} we see that our Ly$\alpha$ escape fractions show a distribution similar to that of the comparison sample. Our inferred Ly$\alpha$ escape fractions resemble the values measured in LARS galaxies, in contrast to those seen in \citet{izo16b} with $f_{\rm esc}$(Ly$\alpha$) values amongst the highest observed in GP galaxies. It is important to mention that the objects studied by \citet{izo16b} were also found to be LCEs. These five LCEs from \citet{izo16b} clearly fall in the upper right quadrant of Figure ~\ref{fig:Halpha_fesc} hinting at a general trend where LCEs have strong Ly$\alpha$ emission, and a very different distribution of EW$(\rm H \alpha$) compared to LARS galaxies, GPs, and those in our sample. \par

\section{Star Formation Properties} \label{sec:SF}
As part of our analysis we estimate SFRs using different methods exploring several wavelength ranges from the UV to the IR. In this section we briefly describe and present our inferred SFR values. 

\subsection{UV Range}\label{sec:UVrange}
We quantify the content of massive stars by comparing the UV spectra to synthetic models \citep{lei13}. As part of the analysis we apply this same technique and use a Starburst99 model \citep{lei99,vaz05,lei09,lei14}. The model parameters are: continuous star formation, age of 20 Myr, Kroupa initial mass function \citep[IMF -][]{kro08} with a mass range of 0.1-100 M$_{\odot}$, solar metallicity (Z= 0.014), spherically extended, blanketed, non-local thermodynamic equilibrium atmospheres, nebular continuum, and stellar evolution models with rotation \citep{eks12,geo13}. We opt for a solar metallicity for our star-forming galaxies given their luminosities and following the mass-metallicity relation of \citet{tre04}. \par
Once the foreground reddening and redshift corrections are applied, we then compare the observed UV spectrum to that of the model to study the intrinsic dust attenuation. For this step we focus on the spectral region covering wavelengths between 1000 and 1400 \r{A}. The wavelength windows affected by geocoronal emission are masked out. We infer the intrinsic reddening, $E(B-V)_{\rm UV}$, by fitting the COS observations ($f_{\rm obs}$) with the unattenuated model ($f_{\rm S99}$ - after scaling the luminosity) reddened by the \citet{cal00} attenuation law, $k(\lambda)$, using the following relationship:
\begin{equation}
f_{\rm S99_{red}}(\lambda) = f_{\rm S99}(\lambda) 10^{-0.4 k(\lambda)E(B-V)_{\rm UV}}.
\end{equation}
The best-fitting $E(B-V)_{\rm UV}$ is determined by iteratively searching for the values that minimize the $\chi ^{2}$,
\begin{equation}
\chi^{2} = \sum \; \frac{(f_{\rm obs} - f_{\rm S99_{red}})^{2}}{\sigma_{\rm err}^{2}}.
\end{equation}
As the Calzetti law does not extend to wavelengths bluer than 1200 \r{A}, we complement the attenuation law with that of \citet{lei02} generated from a sample of star-forming galaxies studied with HUT. The \citeauthor{lei02} relation allows for a smooth transition of the Calzetti law to shorter wavelengths, and their results were later verified by the far-UV attenuation curves derived by \citet{red16} and \citet{bua02}. This addition to the attenuation law facilitates studies near the Lyman break. We use the PYTHON package scipy.optimize to infer the $E(B-V)_{\rm UV}$ values shown in Table ~\ref{tab:SF}. The errors in column 2 are calculated by taking the square root of the covariance matrix of the parameter estimates.\par
This method of estimating the $E(B-V)_{\rm UV}$ is sensitive to the parameters used to compute the model spectrum. In \citet{lei16} we address model assumptions and the impact of varying age, IMF, metallicity, stellar evolution and model atmospheres. The stellar evolution is the parameter that introduces the largest uncertainties, possibly affecting the derived SFRs by up to a factor of two. Furthermore, given the limited wavelength range available for the analysis we do not mask out regions affected by interstellar absorption. We note that interstellar absorption lines might contribute to the overall uncertainties within our inferred  $E(B-V)_{\rm UV}$. \par
The UV luminosity of galaxies is determined by the SFR, which can in turn be easily derived once dust corrections have been taken into account. We correct the observations for intrinsic reddening using the $E(B-V)_{\rm UV}$ values in Table ~\ref{tab:SF}. We take the theoretical monochromatic luminosities from Starburst99 and scale them by 4$\pi D^{2}$ to obtain the unattenuated model spectrum. Using a theoretical model we fit for the best value of the SFR$_{\rm UV}$. Our estimates for the SFR$_{\rm UV}$ of the individual targets are shown in Table ~\ref{tab:SF}. We point out that the SFR$_{\rm UV}$ values derived here are dependent on the centering of the object and vignetting of the PSA caused by extended targets. Given that the COS observations relied on the HST pointing accuracy ($\sim$0.3\arcsec), a small offset in the pointing could lead to variations in the flux. The COS Team, however, states that the flux calibration for extended targets is reasonably accurate for COS FUV observations\footnote{\href{http://www.stsci.edu/hst/cos/documents/newsletters/cos_stis_newsletters/full_stories/2015_03/available_unsupported_modes}{http://www.stsci.edu/hst/cos/documents/newsletters/\\cos\_stis\_newsletters/full\_stories/2015\_03/\\available\_unsupported\_modes}}. Given that the dominant source of measurement uncertainty in the SFR$_{\rm UV}$ is the reddening correction, $E(B-V)_{\rm UV}$, we estimate the errors in the inferred SFR$_{\rm UV}$ values by taking the upper/lower errors in the $E(B-V)_{\rm UV}$ and re-calculating the SFR$_{\rm UV}$. Additionally, in Figure ~\ref{Fig:SpecALLMod} we show the galaxy spectra corrected for redshift, foreground and intrinsic attenuation, along with the best model spectra with the parameters presented in Table ~\ref{tab:SF}. The strong emissions observed in Figure ~\ref{Fig:SpecALLMod} are geocoronal lines (Ly$\alpha$, \ion{N}{I}, and \ion{O}{I}). 

\subsection{Optical Range}\label{sec:optrange}
The zCOSMOS survey \citep{lil09} observed targets in the COSMOS field using the VIMOS spectrograph on the VLT at ESO's Cerro Paranal Observatory, Chile. In the zCOSMOS Data Release DR3 we find optical spectroscopic observations for 6 out of 7 targets studied in this work. We correct the observations for foreground reddening  and redshift as described in Section ~\ref{sec:lyalpha}. \par
We derive the intrinsic color excess, $E(B-V)_{\rm OPT}$, from the Balmer decrements. Similar to the work of \citet{dom13}, in our analysis we assume a value of ($\rm H\alpha/\rm H\beta)_{\rm int} = 2.86$ for Case B recombination corresponding to a temperature $T=10^{4} \rm \;K$ and electron density $n_{e} = 10^{2} \; \rm cm^{-3}$ \citep{ost89}. The color excess is then obtained using the standard relation,
\begin{equation}
E(B-V)_{\rm OPT} = \;1.97 \; \log_{10}\Bigg(\frac{(\rm H\alpha/ \rm H\beta)_{\rm obs}}{2.86}\Bigg).
\end{equation}
We note that ($\rm H\alpha/\rm H\beta)_{\rm obs}$ refer to the ratio of the H$\alpha$ and H$\beta$ emission line fluxes extracted from the foreground-redshift corrected VIMOS observations. No correction for underlying stellar absorption is applied to these values. In column 4 of Table ~\ref{tab:SF} we list the individual $E(B-V)_{\rm OPT}$ values as inferred from the optical observations. Correcting the optical observations for intrinsic reddening using these color excess values we obtain the line fluxes presented in Table ~\ref{tab:Lines}.\par
The observed H$\alpha$ fluxes, $I(\rm H\alpha)$, are transformed to luminosities using the distance ($D$) values described in Section ~\ref{sec:targ_sel} and listed in Table ~\ref{tab:prog}. We estimate the SFR$_{\rm OPT}$ using the inferred $\rm H\alpha$ luminosities, $L(\rm H\alpha)$, along with the linear relation from Starburst99 models
\begin{equation}
\rm SFR_{\rm OPT} = \frac{L(H\alpha)}{3.39 \: \cdot \: 10^{41}\;  \rm erg\; s^{-1}} \rm M_{\odot} \;yr^{-1}.
\end{equation}
The Starburst99 model parameters are similar to those described and adopted in Section \ref{sec:UVrange}. The SFR$_{\rm OPT}$ values derived from the H$\alpha$ luminosities are listed in column 5 of Table ~\ref{tab:SF}. The quoted uncertainties are estimated as described in Section \ref{sec:UVrange}. \par

\subsection{Comparison between UV and Optical}\label{sec:UV_opt_range}
Comparing the average reddening values, $E(B-V)_{\rm UV}$ and $E(B-V)_{\rm OPT}$, we find values of $\sim$0.12 mag and $\sim$0.23 mag, respectively. The difference between the optical depths of the continuum, $E(B-V)_{\rm UV}$, is around one-half of the optical depths of the Balmer emission lines, $E(B-V)_{\rm OPT}$. This is expected, and previously observed in star-forming galaxies by \citet{cal94}. Analyzing the UV and optical spectra of 39 starburst galaxies \citet{cal94} find that the difference between the optical depth in the Balmer emission lines is about a factor of 2 larger than that observed in the continuum underlying the two Balmer lines. These results have been interpreted as the continuum and emission lines sampling different galaxy components of the complete stellar population \citep{kee93}. The ionizing hot stars reside near their birth places, close to dust molecular clouds, therefore their associated nebular recombination radiation is more strongly affected by the presence of dust in the environment. Older stars contributing mostly to the UV and optical continuum, on the other hand, are expected to have drifted away from their progenitor cloud, moving to regions where dust is less concentrated. \par
Turning to the SFRs inferred from the UV/COS observations and those obtained from the optical/VIMOS data we find that the individual SFRs are marginally consistent within the errors, however, there is an offset for most of the targets with the optical SFRs being slightly higher than the UV rates with the exception of 1535411 and 1365128. We find average values of $\sim$0.96 M$_{\odot}$ yr$^{-1}$ and $\sim$2.5 M$_{\odot}$ yr$^{-1}$ for SFR$_{\rm UV}$ and SFR$_{\rm OPT}$, respectively. Looking closer at the individual SFR estimates we find the best agreement in galaxies 1535411 and 1365128, coincidently their SFR$_{\rm UV}$ and $E(B-V)_{\rm UV}$ values are slightly higher than their corresponding
 SFR$_{\rm OPT}$ and $E(B-V)_{\rm OPT}$ estimates. From the values in Table \ref{tab:SF} we identify a trend where the highest offsets between SFR$_{\rm UV}$ and SFR$_{\rm OPT}$ are observed in those galaxies with highest ISM reddening, $E(B-V)_{\rm OPT}$, where SFR appears buried, and is not observed through stellar light. \par
 
To obtain a better picture of the star formation activity in these starburst galaxies we investigate the SFRs using the SED modelling code CIGALE (\citealp{nol09}, Boquien et al., in prep.). The code is based on an energy--balance principle, the energy absorbed by dust from the UV to the near--IR is re--emitted self--consistently in the mid-- and far--IR. To model the galaxies in this sample, we used the COSMOS broadband
observations in the FUV, NUV, u, g, V, r, i, z, Spitzer/IRAC 3.6~$\mu$m,
Spitzer MIPS 24~$\mu$m, Herschel/PACS 100~$\mu$m and 160~$\mu$m, and
Herschel/SPIRE 250~$\mu$m, 350~$\mu$m, and 500~$\mu$m.
We find an average SFR value of $\sim$ 6 $\pm$ 1 M$_{\odot}$ yr$^{-1}$, higher than the inferred SFRs from the UV and the optical. This indicates there is more star formation activity buried than what is seen in the UV and optical ranges.  \par
We also note that the VIMOS observations use 1.0\arcsec\: slits, which do not enclose the galaxies entirely. Given the different morphologies, a small offset in the pointing could lead to variations in the flux. In general, the SFR$_{\rm OPT}$ values might better represent a lower limit on the true SFR. \par 
One explanation for the difference between the SFRs can be attributed to attenuation caused by dust. Adopting the dust luminosity computed by CIGALE and combining it with the dust SFR estimator of \citet{ken12}, we find SFRs that are more comparable to initial CIGALE SFR values than those inferred from the UV and optical observations. Care is required when comparing the different values presented in this work as the inferred SFRs have an intrinsic dependance on the models used. As mentioned in Section \ref{sec:UVrange} for the case of the SFR$_{\rm UV}$, the stellar evolution used in estimating the $E(B-V)_{\rm UV}$ introduces the largest uncertainties possibly impacting the derived SFRs by a factor of two. \par
Given the main scope of this paper we will not further discuss the possible sources of uncertainty in the SFRs presented here, and will simply note that an in-depth investigation is needed to fully understand the complexity of such measurements. \par

\begin{table*}
\caption{Star Formation Properties }
\label{tab:SF}
\centering
\begin{tabular}{ccccc}
 \hline  \hline \
& \multicolumn{2}{c}{UV/COS}& \multicolumn{2}{c}{Optical/VIMOS}\\
Target & $E(B-V)_{\rm UV}$ & SFR$_{\rm UV}$ & $E(B-V)_{\rm OPT}$ & SFR$_{\rm OPT}$ \\
 & (mag) & ($M_{\odot} \rm \: yr^{-1}$)& (mag) & ($M_{\odot} \rm \: yr^{-1}$)  \\
  \hline\\ 
1084255& 0.20$\pm$0.08 & 1.51$^{+2.32}_{-0.92}$&0.53$\pm$0.18& 5.4$^{+5.1}_{-2.0}$ \\
1535411&	0.22$\pm$0.16& 2.18$^{+13.73}_{-1.88}$&0.06$\pm$0.02& 1.4$^{+0.1}_{-0.1}$ \\
1511408&	0.05$\pm$0.15& 0.18$^{+0.87}_{-0.08}$&0.21$\pm$0.08& 1.0$^{+0.3}_{-0.2}$ \\
1508056&	0.05$\pm$0.14& 0.31$^{+1.48}_{-0.15}$&0.14$\pm$0.02& 1.2$^{+0.1}_{-0.1}$ \\
1235867&	0.09$\pm$0.14& 0.21$^{+0.78}_{-0.14}$& - & - \\
781126& 0.07$\pm$0.02& 1.48$^{+0.43}_{-0.29}$&0.33$\pm$0.13& 5.4$^{+2.8}_{-1.6}$  \\
1365128&	0.13$\pm$0.07& 0.84$^{+1.04}_{-0.48}$&0.11$\pm$0.10& 0.6$^{+0.3}_{-0.1}$ \\
 \hline 
 \end{tabular}
\end{table*}

\begin{table*}
\caption{Emission Lines}
 \label{tab:Lines}
\centering
\begin{tabular}{cccccccccc}
 \hline  \hline \
Target & $I(\rm L$y$\alpha$) $^a$ &
$EW(\rm L$y$\alpha$)$^b$&
$I(\rm H\alpha$)$^a$&
$EW(\rm H\alpha$)$^b$&
$I(\rm H\beta$)$^a$&
$EW(\rm H\beta$)$^b$&
$I(\ion{O}{III}$)$^a$&
$L(\rm H\alpha)$$^c$ &
$f_{\rm esc}$(Ly$\alpha$)$^d$\\ 
  \hline\\ 
1084255& 69.4&33&79.3& 66& 30.6&9&35.6& 18.3& 10.1\\
1535411&	-&-&13.7 & 452&4.8&77&16.2&4.7& -\\
1511408&-&-&14.5 & 38 &5.3&7&4.7&3.2& -\\
1508056&	6.1&147&11.7&210&4.1&27&15.7& 4.2& 6.0\\
1235867&	-&-&- &- &-&- & -& -&-\\\
781126& 36.94& 18&77.3& 226& 28.0&36&126.0&18.1& 5.5\\
1365128&	9.6&98.0 &8.0&34&3.1&7& 4.6&1.9& 13.0\\
 \hline 
 \end{tabular}
 \\ 
 \begin{minipage}{16cm}~\\
 \textsuperscript{$a$} Observed flux density in 10$^{-16}$ erg s$^{-1}$ cm$^{-2}$. These values include foreground and intrinsic dust corrections. $I(\rm Ly\alpha$) have been corrected for stellar absorption.\\
\textsuperscript{$b$}{Restframe equivalent width in \r{A}. $EW$(Ly$\alpha$) have been corrected for stellar absorption.}\\
\textsuperscript{$c$}{Luminosity in 10$^{41}$ erg s$^{-1}$.}\\
\textsuperscript{$d$}{Escape fraction of Ly$\alpha$ in \%.}\\
\end{minipage}
\end{table*}

\section{Lyman Continuum Escape Fractions} \label{sec:esc}
We now focus on the COS observations to examine the Lyman continuum (LyC) emission of our targets. With a  minimum redshift of $z\sim 0.26$ and using the COS G140L/1105 mode we observe uncontaminated LyC ($\lambda_{\rm rest}<912$ \r{A}) in a window of $\sim$30 \r{A}. \par
In order to measure the escape fraction of LyC photons, $f_{\rm esc}(\rm LyC)$, we measure the expected LyC emission at $\sim$900 \r{A} from the modelled flux, produced by the massive stars, and compare it to the observed fluxes. A comparison of the predicted fluxes with the dereddened observed flux yields an estimate of the relative escape fraction, $f_{\rm rel}$. This escape fraction refers to the percentage of LyC photons that would escape the galaxy in the absence of dust. To estimate the absolute escape fraction, $f_{\rm abs}$, instead we compare the modelled flux to those of the observations prior to the intrinsic reddening correction. This $f_{\rm abs}$ accounts for the presence of both, gas and dust as seen in the UV range. We note that as stated in Section \ref{sec:UV_opt_range} there is a possibility that some of the star formation activity may be hidden and unaccounted for by the dust observed in the UV wavelength regime. \par

Given the low flux level of the observations we apply a weighted stacking approach to the galaxy spectra, excluding the QSO and masking those wavelength regions contaminated by geocoronal emission. We define the weights as $w$ = 1/$\sigma_{\rm err}$, where $\sigma_{\rm err}$ are the statistical errors (i.e. Poisson noise). We obtain two combined sets, dereddened (corrected for both Milky Way and intrinsic reddening), and observed (corrected only for Milky Way reddening) spectrum. Before combining the individual spectra into a single one we normalize the observations by the median of the flux between restframe 1220 $\leq \lambda \leq$ 1230 \r{A}. In Figure ~\ref{Fig:CombSpec} we present the combined spectrum. Note that the strong emission around 1030 \r{A} is the residual of geocoronal contamination.\par

 The escape fractions inferred as part of this work are dependent on the model atmospheres against which we compare our observed LyC. We make use of the interface WM-Basic \citep{pau98} to model the atmospheres of O stars, mainly optimized for the UV spectral range. WM-Basic implemented a simplified treatment of hydrogen line modelling since there are no valuable diagnostic hydrogen lines in this particular range. More specifically, WM-Basic does not account for Stark broadening, causing the Lyman lines originating from the photosphere to be relatively weak in the models. Although this simplification in the models would typically affect all of the Lyman lines beyond Lyman-$\beta$, \citet{kas12} showed that the atmospheric structure is generally not affected by this treatment. As detailed in \citet{lei16} the only wavelengths compromised by the omission of Stark broadening are those between 912-920 \r{A}. In this region we observe an artificial rise in the continuum caused by the lack of Lyman blanketing, therefore we exclude these wavelengths when determining the Lyman break.

We compare the data below 912 \r{A} to the theoretically predicted flux shown in yellow in Figure ~\ref{Fig:CombSpec}. We average the flux over an interval of 18 \r{A} between 892 to 910 \r{A} for all three sets, the theoretically predicted SED ($F_{\rm \lambda,mod}$), the reddened corrected spectrum ($F_{\rm \lambda,dered}$), and the uncorrected spectrum ($F_{\rm \lambda,obs}$). This wavelength window was mainly chosen to avoid regions affected by the noisy detector edges. The LyC escape fractions are inferred from the ratios $\frac{F_{\rm \lambda,dered}}{F_{\rm \lambda,mod}}$ and $\frac{F_{\rm \lambda,obs}}{F_{\rm \lambda,mod}}$ for $f_{\rm rel}$ and $f_{\rm abs}$, respectively. In Table ~\ref{tab:fesc} we present our measured LyC escape fractions. \par
From the dereddened observations we find a relative escape fraction of 1.7$^{+15.2}_{-1.7}$\%. This value represents the percentage of LyC photons that would escape such galaxies in the absence of dust. A more representative estimate of the escaping LyC photons is the absolute escape fraction, $f_{\rm abs}$, which accounts for the obscuration caused by the intrinsic dust. We infer an absolute escape fraction of  0.4$^{+10.1}_{-0.4}$\%. The uncertainties account for the statistical errors extracted and propagated from the x1d files ($\sigma_{i}^{2}$). The final errors listed in Table ~\ref{tab:fesc} were computed from the combined spectra as follows

\begin{equation}
\overline{\sigma} = \sqrt{ \frac{\sum_{}^{} \sigma_{i}^{2}}{N^{2}}}.
\end{equation}

Given the relatively large uncertainties in the derived escape fractions we conservatively consider these values as upper limits. In general a low absolute escape fraction agrees with the predictions from \citet{ver15} proposing that when present, the Ly$\alpha$ peak separation traces the column density of the scattering medium. They predict that peak separations in the order of $\gtrsim\:$300 km s$^{-1}$ would describe objects in the optically thick regime where LyC photons would be unable to escape. We point out that we measure peak separations in the order of  $\gtrsim\:$300 km s$^{-1}$, where one would expect no escape of LyC radiation due to the thick ISM. \par

\citet{izo16a}, \citet{izo16b} and \citet{izo18} have measured some of the highest LyC escape fractions in five compact star-forming galaxies in low-redshift galaxies. For these same galaxies \citeauthor{izo16a} measure Ly$\alpha$ escape fractions ranging between 22-98\%. Compared to the $f_{\rm esc}(\rm Ly\alpha)$ from \citeauthor{izo16a}, we measure escape fractions of Ly$\alpha$ ranging from $\sim$5 to 13\%. The low LyC escape fraction we infer is also supported by the relatively low Ly$\alpha$ escape fractions measured from the individual galaxies. \citet{ver17} show that a clear correlation between $f_{\rm esc}(\rm Ly\alpha)$ and $f_{\rm esc}(\rm LyC)$ exists. However, LCEs relevant to the cosmic reionization, $f_{\rm esc}(\rm LyC)>$10 \%, exhibit $f_{\rm esc}(\rm Ly\alpha)$ with values $>$20 \%. \par

For comparison we also combine the corrected (redshift, foreground, and intrinsic reddening) VIMOS observations using the same stacking procedure described above. From this optical combined spectrum we then measure the Ly$\alpha$ and H$\alpha$ fluxes and following the recipe described in Section \ref{sec:lyalpha} we infer a relative low escape fraction of  $f_{\rm esc}(\rm Ly\alpha)$ = 2.6 \% (also listed in Table \ref{tab:fesc}). Caution is required when comparing this value to the $f_{\rm esc}(\rm LyC)$ limits above as the optical combined spectra includes only six out of the seven objects combined in the UV spectrum. \par

   \begin{figure*}
            {\includegraphics[scale=0.65]{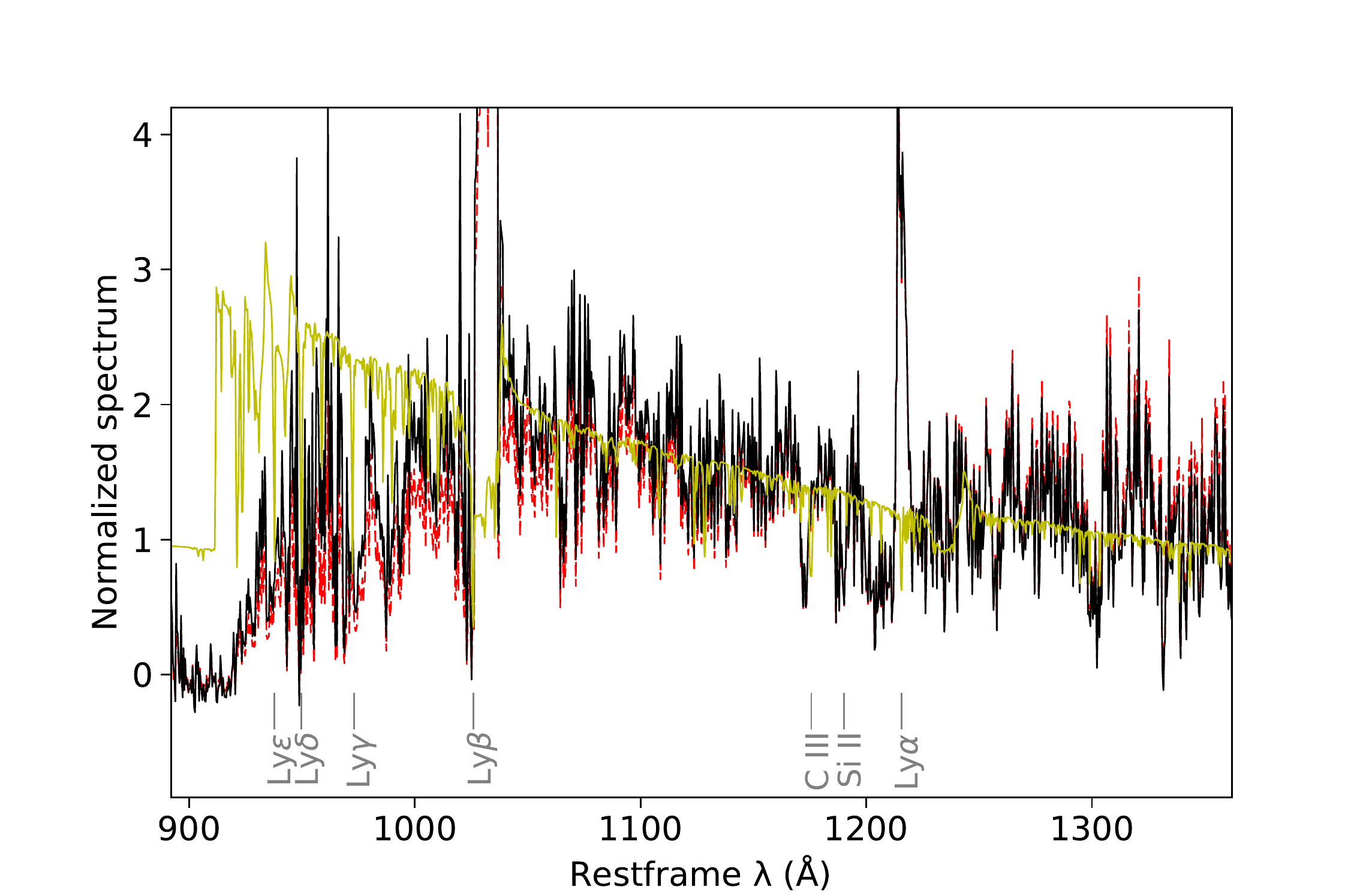}}
            \centering
      \caption{COS combined observations. In black we show the dereddened spectrum, in red the reddened/uncorrected spectrum, and in yellow the model SED. At the  bottom of the panel we identify intrinsic spectral lines. }
         \label{Fig:CombSpec}
   \end{figure*}
   
   \begin{figure}
            {\includegraphics[scale=0.7]{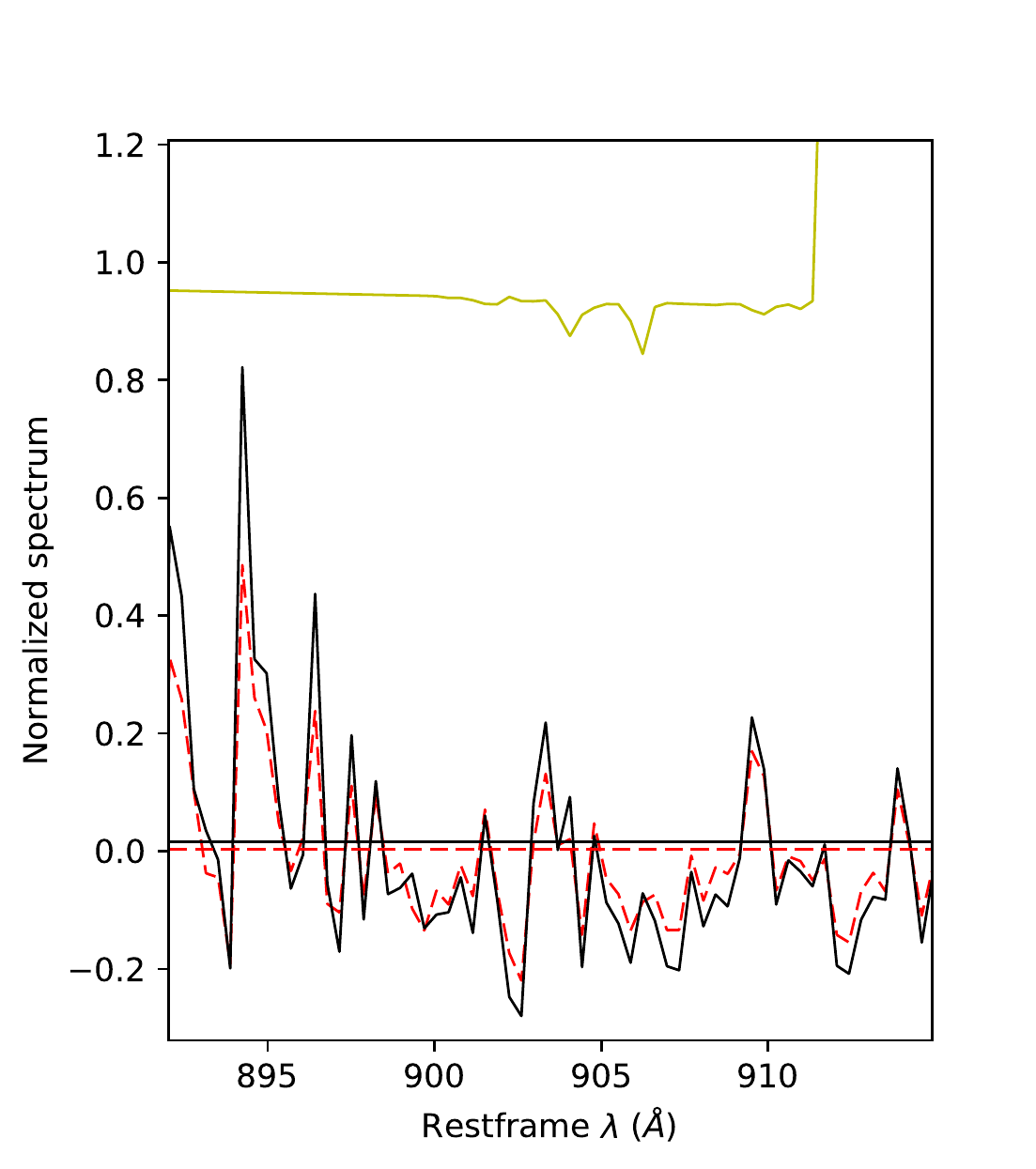}}
            \centering
      \caption{Spectral region around the Lyman break. We show the combined dereddened spectrum in black, and in red dashed line the observed spectrum. We present with horizontal lines the inferred mean flux densities, in solid black the mean flux for the dereddened observations, and in red dashed line the mean flux for the observed spectrum. In yellow we display the modelled flux. Between wavelengths 892-910 \r{A} there are $\sim$ 40 resels.}
         \label{Fig:CombSpec}
   \end{figure}
   
\begin{table}
\caption{Lyman escape fractions of combined observations}  
\label{tab:fesc}
\centering
\begin{tabular}{ccc}
 \hline  \hline \
$f_{\rm rel}$(LyC)& $f_{\rm abs}$(LyC) & $f_{\rm esc}$(Ly$\alpha)$\\
(\%) & (\%) & (\%) \\
  \hline\\ 
1.7$^{+15.2}_{-1.7}$& 0.4$^{+10.1}_{-0.4}$& 2.6\\
 \hline 
 \end{tabular}
\end{table}

\section{Conclusions} \label{sec:conclusion}
In this paper we present the analysis of seven star-forming galaxies observed with HST/COS at redshifts $z\sim0.3$. We aim to study the escape of LyC and Ly$\alpha$ radiation from these galaxies. Our main findings can be summarized as follows:

\begin{itemize}
\item We observe Ly$\alpha$ emission in 4 out of the 7 galaxies studied in detailed in this work. 
\item We find double peak features in the Ly$\alpha$ profiles from 1084255, 1365128 and 1508056, with corresponding peak separations of 655, 374, and 275 km s$^{-1}$. The relatively higher peak separation values measured for 1084255 and 1365128 might hint at an absence of leaking Lyman continuum photons \citep{ver15}. 
\item 3 out of the 4 objects with Ly$\alpha$ emission have EW(Ly$\alpha$) > 20 \r{A}. Galaxies with such EW values are classified as Ly$\alpha$ emitters (LAE) in high-$z$ narrow-band surveys. 
\item We estimate the Ly$\alpha$ escape fraction, $f_{\rm esc}(\rm Ly\alpha)$, and find values up to 13\%. We see that our measurements follow the correlation between Ly$\alpha$ EWs and the $f_{\rm esc}(\rm Ly\alpha)$ observed by \citet{yan17} in GP galaxies. 
\item After combining the individual galaxy spectra we conservatively infer upper limits in the absolute escape fraction of the order of $<0.4^{+10.1}_{-0.4}$ \%, concluding that these galaxies are optically thick to Lyman continuum radiation. This demonstrates the validity of H$\alpha$ as a star-formation tracer in these galaxies. 
\item Comparing several star-formation tracers, we find that the inferred SFRs are strongly dependent on the model assumptions. 
\item We find moderate SFR values for most of the galaxies studied in this work (SFRs $\lesssim$ 10 M$_{\odot}$ yr$^{-1}$). 
\end{itemize}

\section*{Acknowledgements}

The work here was based on observations from Program ID 13313 made with the NASA/ESA \textit{Hubble Space Telescope}, obtained from the data archive at the Space Telescope Science Institute (STScI). Based on zCOSMOS observations carried out using the Very Large Telescope at the ESO Paranal Observatory under Programme ID:LP175.A-0839.  Support for this work has been provided by NASA through grant No. GO-13313 from the Space Telescope Science Institute, which is operated by AURA, Inc., under NASA contract NAS5-26555. MB was supported by the FONDECYT regular project 1170618 and the MINEDUC-UA projects codes ANT 1655 and ANT 1656.








\appendix
\clearpage

\section{COS NUV Images}\label{appendix_nuv}
\begin{figure*}
            {\includegraphics[scale=0.27]{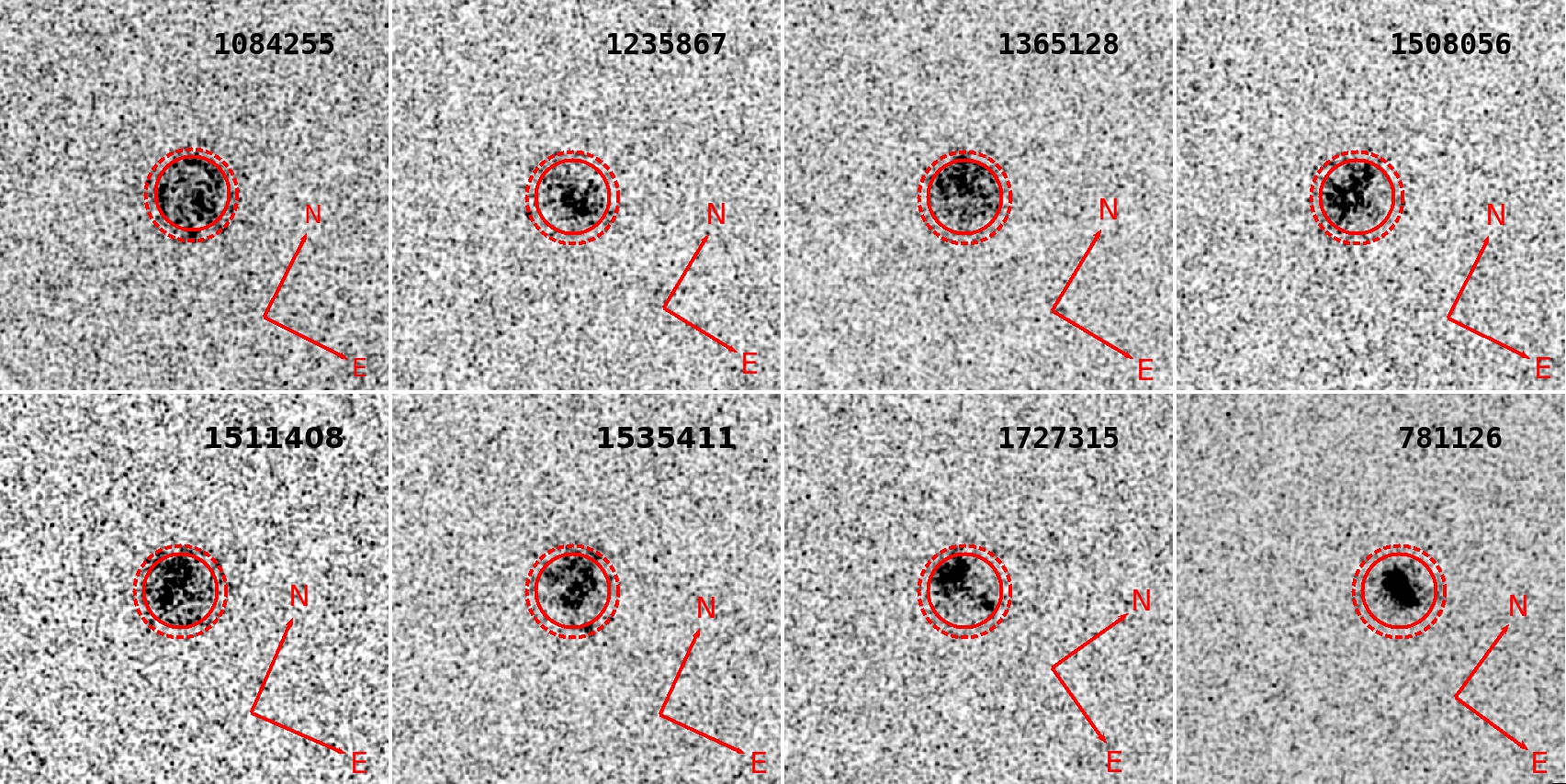}}
            \centering
      \caption{13\arcsec $\times$ 13\arcsec HST/COS NUV Images using MIRRORA. The frames were taken as part of the HST program 13313. Solid red circles shows the expected location of the COS 2.5\arcsec aperture. The dashed red circles show the HST pointing error estimated to be $\sim$0.3\arcsec. Each stamp shows the orientation of the image.}
         \label{fig:nuv}
   \end{figure*}

As described in Section \ref{sec:obs_data}, HST program 13313 did not perform target acquisitions before observing the objects, instead the program relies on the typical HST pointing accuracy estimated to be $\sim$0.3\arcsec. Although no acquisition was performed, short exposures ($t_{\rm exp} = 120$s) were taken in imaging mode using the NUV MIRRORA element. Before proceeding with the analysis presented here we inspect the individual NUV images to confirm the galaxies were indeed in the field of view, on the detector, and inside the COS PSA. In Figure \ref{fig:nuv} we show the individual NUV frames. We show the expected location of the COS PSA aperture as solid red circles, and represent the HST pointing accuracy with dashed red circles. We note that the exact location of the PSA on the NUV images is not known, however, with an estimated pointing error of $\sim$0.3\arcsec\: as shown in Figure \ref{fig:nuv} the targets are still contained inside the COS aperture. \par
\begin{figure}
            {\includegraphics[scale=0.39]{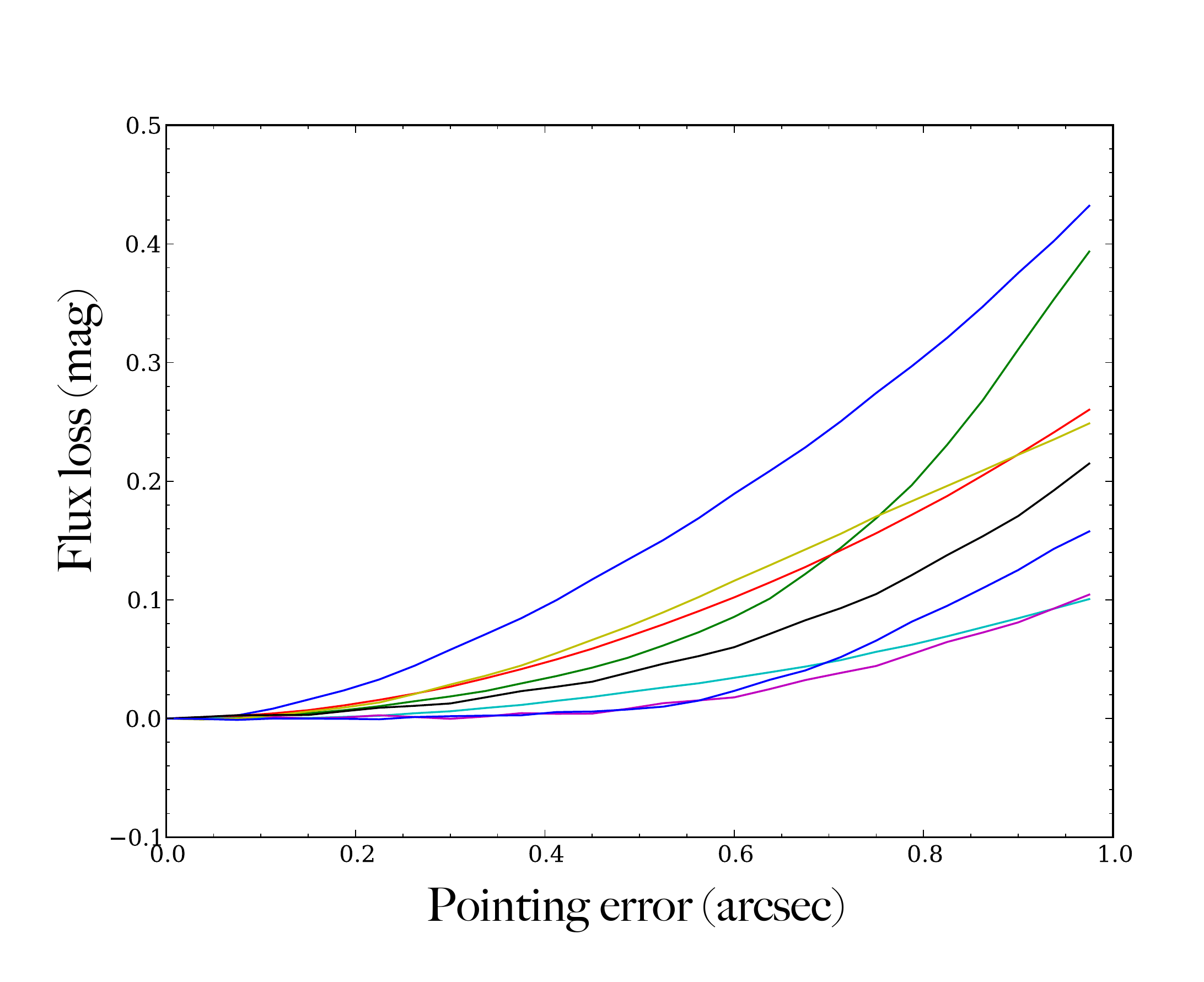}}
            \centering
      \caption{Estimated target flux loss due to pointing offset as a function of offset. Each of the curves represents a different target observed in HST program 13313. The flux loss is estimated by averaging the pointing offsets in various directions. With an offset representative of the the HST pointing accuracy ($\sim$0.3\arcsec), the typical flux loss is estimated to be $\lesssim$0.05 mag. }
         \label{fig:offset}
   \end{figure}

Furthermore, looking at the flux loss due to different pointing offsets (Figure \ref{fig:offset}) we find that for an offset of 0.3\arcsec\: the typical flux loss is estimated to be around $\lesssim$0.05 mag. Even for extreme cases where the observations experience offsets of the order of 0.7\arcsec, rare in HST observations, the flux loss is limited to $\sim$0.15 mag.

\section{Updates to CalCOS v2.21d}\label{appendix_calcos}
In \citet{lei16} we introduced a modified version of CalCOS to perform a 2D background correction available for data taken at lifetime position 1 (LP1). In the modified version of CalCOS, v2.21d, anytime the DARKFILE keyword is found in the primary header of the science exposures the software subtracts the background contribution using a two-dimensional superdark. The user must add manually the DARKFILE keyword and specify the name of the superdark reference file. The dark-subtracted FLT is stored in files with extensions $darkcorrflt\_a/b.fits$.\par

In CalCOS v2.21d, before performing the spectral extraction we implement a weighting system that assigns low weights to those pixels in the science frames that did not register any counts. After the 2D background correction, these zero-count pixels end up with negative values. Given the low weights assigned to these pixels, the flux extracted from these pixels is close to zero, in the order of $\sim10^{-21}$ ergs s$^{-1}$ cm$^{-2}$ \r{A}$^{-1}$. In order to avoid any biases in the average extracted flux towards higher values, we modify CalCOS v2.21d to remove the weighting system, and instead we now assign a similar weight for all pixels irrespective of positive or negative values. This change is especially critical for those regions on the detector with low sensitivity. \par
\begin{figure}
            {\includegraphics[scale=0.45]{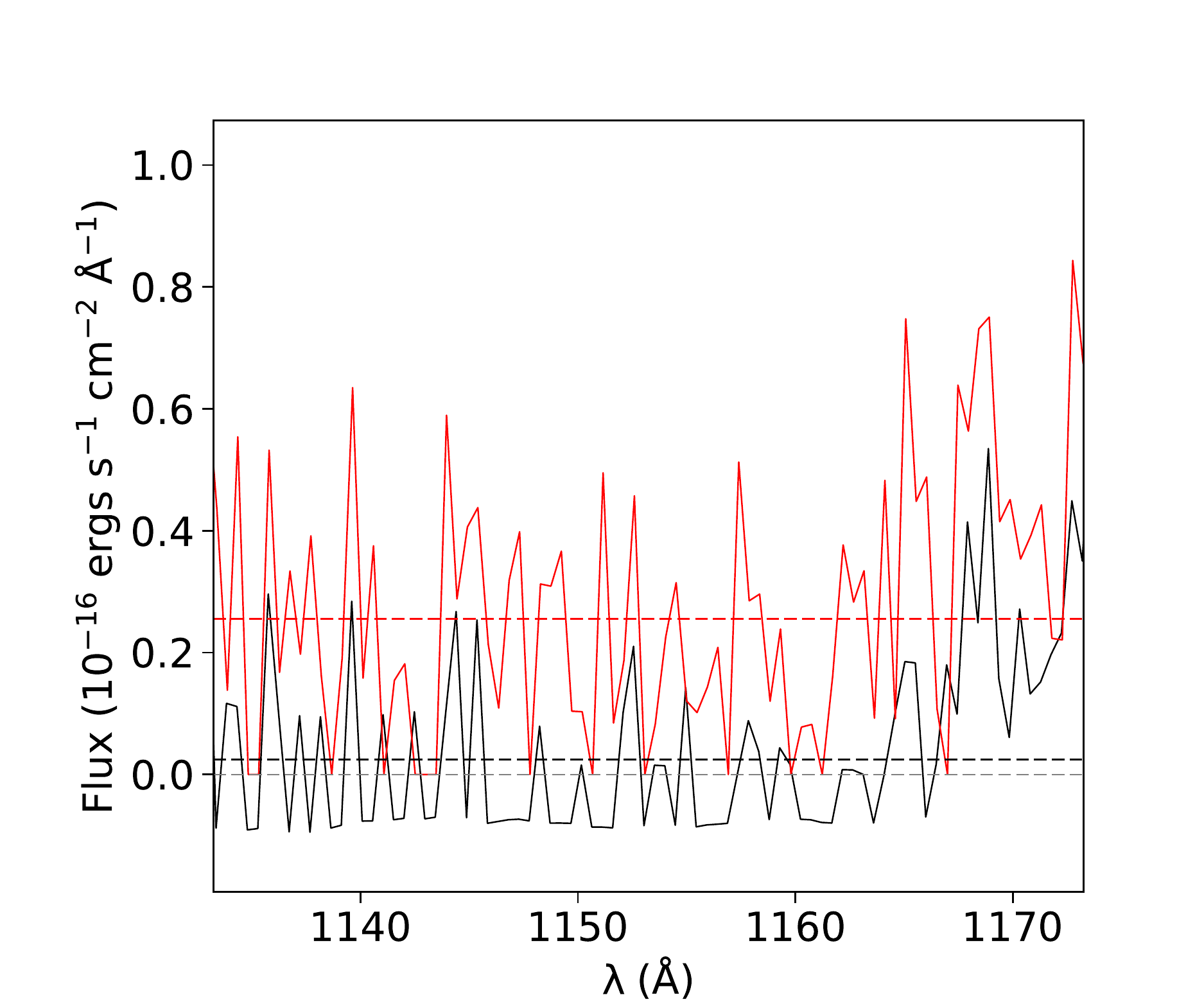}}
            \centering
      \caption{Science observations for 781126, flux as a function of observed wavelength. In black we show the data calibrated with the modified version of CalCOS v2.21d, which includes uniform weights. In red we display the spectrum calibrated with CalCOS v2.21d where pixels with zero counts are assigned low weights. The black and red dashed lines show the mean flux values for wavelengths 1135 \r{A} $<\lambda<$ 1170 \r{A}. Zero fluxes are marked with a grey dashed line. The spectra has been binned to a COS resolution element (1 resel = 6 pixels).}
         \label{Fig:app}
   \end{figure}

In Figure ~\ref{Fig:app} we compare the calibrated spectra from CalCOS v2.21d (in red) and from the modified software (in black). We see that the changes made to CalCOS v2.21d decreases the mean flux $<$10 $\%$ between observed wavelengths 1135 \r{A} $<\lambda<$ 1170 \r{A}. This is mainly due to the fact that by assigning low weights to zero-count pixels, CalCOS v2.21d avoided negative fluxes and instead provided extremely low flux values. With the new modifications to CalCOS v2.21d we now allow for negative values, as expected from uncertainties in the superdarks. We point out that the floor at negative fluxes seen in the black spectrum in Figure ~\ref{Fig:app} is caused by the background subtraction and the decreasing instrument sensitivity. 

\bsp	
\label{lastpage}
\end{document}